\pdfoutput=1
\documentclass{jfm}
\usepackage{graphicx}
\usepackage{epstopdf, epsfig}
\usepackage{url}
\usepackage{dcolumn}
\usepackage{bm}
\usepackage{amsmath}
\usepackage{amsfonts}
\usepackage{subfigure}
\usepackage{xcolor}

\newcommand{\be}{\begin{equation}}
\newcommand{\ee}{\end{equation}}
\newcommand{\bea}{\begin{eqnarray}}
\newcommand{\eea}{\end{eqnarray}}

\newcommand{\llangle}{\left\langle}
\newcommand{\rrangle}{\right\rangle}
\newcommand{\avg}[1]{\llangle #1 \rrangle}

\DeclareMathSymbol{\shortminus}{\mathbin}{AMSa}{"39}
\shorttitle{Critical transition to a non-chaotic regime in isotropic turbulence}
\shortauthor{D. Clark, A. Armua, R.D.J.G Ho and A. Berera}

\title{Critical transition to a non-chaotic regime in isotropic turbulence}

\author{Daniel Clark\aff{1}
  \corresp{\email{daniel-clark@ed.ac.uk}}, Andres Armua\aff{1},
  Richard D. J. G. Ho\aff{2}
 \and Arjun Berera \aff{1}}

\affiliation{\aff{1}SUPA, School of Physics and Astronomy, University of Edinburgh, JCMB, \\ King’s Buildings, Peter Guthrie Tait Road EH9 3FD, Edinburgh, United Kingdom.
\aff{2}Institute of Theoretical Physics, Jagiellonian University, Łojasiewicza 11, 30-348, Kraków, Poland}

\begin{document}

\maketitle

\begin{abstract}
We study the properties of homogeneous and isotropic turbulence in higher spatial dimensions through the lens of chaos and predictability using numerical simulations. We employ both direct numerical simulations (DNS) and numerical calculations of the Eddy Damped Quasi Normal Markovian (EDQNM) closure approximation. Our closure results show a remarkable transition to a non-chaotic regime above critical dimension $d_c\approx 5.88$. We relate these results to the properties of the energy cascade as a function of spatial dimension in the context of the idea of a critical dimension for turbulence where Kolmogorov's 1941 theory becomes exact.
\end{abstract}

\section{Introduction}
\label{se:intro}

The study of the fundamental properties of fluid turbulence have come to be dominated by the 1941 theory of \citet{kolmogorov1941local} (K41) and the arguments surrounding its validity. One of the main predictions of this theory is that, for turbulence at asymptotically high Reynolds numbers ($\Rey$) that is homogeneous and locally isotropic, there should exist a range of length scales over which the statistical properties are determined solely by the mean rate of energy dissipation $\avg{\varepsilon}$. This has come to be known as the inertial range in which K41 theory predicts a multi-step cascade of energy from large to small scales as large structures in the flow break down into progressively smaller structures. By coupling the K41 hypotheses with dimensional analysis, it is straightforward to show that in the inertial range the energy spectrum, $E(k)$ takes the form \begin{equation}\label{eq: k41}
    E(k) = C \avg{\varepsilon}^{\frac{2}{3}}k^{-\frac{5}{3}},
\end{equation} where $C$ is a universal constant.  For statistically stationary turbulence the mean rate of energy dissipation at the smallest scales will be equal to the mean rate of energy injection at the largest scales. The energy cascade of the inertial range must then be driven by a mean energy transfer rate $\avg{\Pi}$, which numerically must be equal to $\avg{\varepsilon}$. As such, it is often argued that this mean energy transfer rate, being an inertial range quantity, should replace the mean dissipation rate in the inertial range energy spectrum \citep{kraichnan_1974, bowman_1996, mccomb2009scale}. Note, that the relationship between $\avg{\Pi}$ and $\avg{\varepsilon}$ is purely kinematic in nature, the underlying mechanisms behind inertial energy transfer and energy dissipation are different, hence on physical grounds this distinction is important, though often overlooked.

The K41 theory of turbulence is appealing in its simplicity, however, its validity remains an open, and somewhat controversial, question. In 1962, motivated by a criticism from Landau \citep{LandauFluid}, Kolmogorov updated his theory to account for what has become known as internal intermittency in his K62 theory \citep{kolmogorov1962refinement}. This new theory introduced an anomalous exponent to the energy spectrum as well as to the real space structure functions. Initial experimental measurements of these structure functions, for example in \citep{van1970structure}, found that at fourth order the exponent $\zeta_4$ did not agree with the K41 prediction and was in better agreement with the K62 value. However, things were further complicated by the more detailed experimental study of \citet{anselmet1984high} showing that at higher orders of the structure functions neither K41 nor K62 values were seen for the exponent. Attempts to reconcile theory with the experimental results led to the development of numerous fractal models \citep{frisch1978simple,frisch1980fully}. These ideas of multi-scaling in turbulence have been well explored in recent years \citep{schumacher2007asymptotic, yakhot2017emergence, sinhuber2017dissipative, iyer2020scaling} with good agreement with the theory found.

In the development of K62 and later theories discussed above, there is an explicit assumption that anomalous scaling in turbulence is a result of internal intermittency, and thus as K41 does not account for this, it must be incorrect. However, \citet{kraichnan_1974, kraichnan1991turbulent} highlighted that K41 cannot be ruled out solely due to spatial variation in the energy dissipation rate. Hence, another viewpoint that claims deviations from K41 scaling are due to finite Reynolds number effects has emerged. This idea was discussed in detail by \citet{jian1994skewness, qian1997inertial, qian1998normal, qian1999slow} and in recent years, a substantial body of work in this area has formed, both experimental and numerical \citep{mccomb2014spectral, tang2017finite, tang2018reappraisal, tang2019can, meldi2018reynolds, djenidi2019scale, antonia2017small, antonia2019finite}.

While corrections to K41 due to either intermittency or finite Reynolds numbers have received substantial and sustained interest, we consider now a third possibility that has been explored periodically over the last half century. The K41 inertial range energy spectrum exhibits scale invariance which has drawn many to make comparisons between fluid turbulence and critical phenomena \citep{nelkin1974turbulence, nelkin1975scaling,de1975phase, rose1978fully, bramwell1998universality, aji2001fluctuations, yakhot2001mean, giuliani2002critical, verma2004statistical, frisch2012turbulence}. These analogies are interesting when considering the concept of a critical spatial dimension found in critical phenomena \citep{ginzburg1960some, wilson1972critical}. Above this dimensions the predictions for the critical exponents of the system are given by mean field theory. With its use of the mean energy transfer rate, it has been suggested that K41 is in fact a mean field theory, exact only above a critical dimension \citep{siggia1977origin, bell1978time}.

Understanding the properties of the turbulent energy cascade predicted by K41 is of great importance if the theory is to be validated or rejected. One method of extracting information about the energy cascade is to vary the spatial dimension of the system and observe the effect. Most famously, in two dimensions the direction of the energy cascade is reversed, and energy flows from small to large scales. Such behaviour was predicted independently by Kraichnan, Leith and Batchelor \citep{kraichnan1967inertial,leith1968diffusion,batchelor1969computation}. This inverse cascade can be understood as a consequence of an additional positive-definite ideal invariant, the enstrophy, in two dimensions. 

Recent studies into turbulence in dimensions higher than three \citep{gotoh2007statistical, yamamoto2012local, berera2020homogeneous, clark2021effect} have shown far smaller differences when compared with three dimensions. However, some changes related to the energy cascade were observed. Notably, in \citet{berera2020homogeneous} a dramatic reduction in fluctuations of the total energy was observed. Indeed this reduction was far larger than could be expected from central limit theorem argument. Additionally in \citep{clark2021effect} using the eddy damped quasi-normal markovian (EDQNM) approximation it was found that with increasing dimension the energy spectrum bottleneck effect \citep{falkovich1994bottleneck} is enhanced. Furthermore, it was found that the velocity derivative skewness increased until around eight dimensions before decreasing towards a potentially dimension independent constant. Moreover, related to the velocity derivative skewness it was found that the enstrophy production reached a maximum value near six dimensions.

In this work, we are motivated by this idea of a critical dimension for turbulence.  However, in this work we turn our attention  to the chaotic properties of turbulence as the spatial dimension is varied through measurement of the maximal Lyapunov exponent. The chaotic properties of turbulence have been shown to be highly dependent on the spatial dimension for the two- and three-dimensional cases \citep{ Berera2018,clark2020chaos, clark2020chaotic}. As such, we may expect to see some influence on the chaotic properties based on the effects seen as the dimension is increased. Additionally, any changes to the chaotic properties may help shed light on the cascading behaviour in higher dimensions.

Clearly, experimental flows are restricted to three dimensions, hence we must make use of numerical and analytical methods. To this end, we employ direct numerical simulation, building on our work in \citep{berera2020homogeneous}, and the EDQNM approximation, following on from \citep{clark2021effect}. As in \citep{clark2021effect}, where appropriate we consider comparisons between DNS and EDQNM results. 

The remainder of this paper is organised as follows: in section \ref{se:ddim} we review and critically appraise the research into turbulence as a function of spatial dimension carried out to date. Next, in section \ref{se:chaosInt} we discuss predictability in turbulent flows and introduce the approach and methodology used in this study. In section \ref{se:results}  we present the behaviour of the error growth between three and eight spatial dimensions, including  a comparison with DNS results in $d=3$ and $4$. Here we find a critical dimension for error growth at $d\approx5.8$. Finally, in section \ref{se:conclusions} we discuss the implications of our results.

\section{Turbulence as a function of spatial dimension}\label{se:ddim}
\subsection{Crossover dimensions}

As we briefly mentioned in the previous section, in a series of works in the early 1970's by \citet{nelkin1974turbulence,nelkin1975scaling} an analogy between fluctuations in turbulence at high $\Rey$ and critical-point fluctuations in critical phenomena was made. These gave a prediction of the scaling exponents of turbulence and studied the existence of a crossover dimension, below which the Kolmogorov scaling relations in K41 become exact \citep{kolmogorov1941local}, in analogy with mean field theory becoming exact above the crossover dimension $d=4$ for the Ising model \citep{wilson1972critical}. The existence of such crossover dimension was criticised later in \citep{frisch1976crossover}. Based on Nelkin's ideas, Kraichnan looked at the statistical properties of a passive scalar field advected by an incompressible $d$-dimensional turbulent flow, finding a strong dependence on dimensionality and a suppression of temporal fluctuations in the stretching rates at $d \rightarrow \infty $ \citep{kraichnan1974convection}. Subsequent numerical work in this direction was done for incompressible flows in \citep{gat1998anomalous,mazzino2000passive} and further generalizations to compressible flows in \citep{gawedzki2000phase,celani2010dispersion}. The idea of crossover dimensions in fully developed turbulence also led to the use of dynamic renormalization group methods \citep{forster1976long,forster1977large,dedominicis1979energy}.

The existence of a critical dimension between two and three dimensions was further investigated in \citep{frisch1976crossover,fournier1978d} by considering non-integer dimensions such that the conservation laws are weakly broken. In these works, an EDQNM approximation is used to show that for $d\gtrsim 2$ the direction of the cascade reverses. A different approach is to modify the aspect ratio of the lattice from cubic $d=3$ to $d=2$ observing the same cascade reverse \citep{benavides2017critical}. The second of these is a more realistic representation of atmospheric turbulence, however, it is less suited for analytic investigation. The third approach to non-integer dimensions can be found in \citep{frisch2012turbulence, lanotte2015turbulence} where a fractal decimation in Fourier space is used to reduce the effective dimensionality of the system in DNS. The first of these works studied effective non-integer dimensions below $d=2$, finding a critical dimension at $d=4/3$ where the energy flux vanishes. Whilst in the second, starting from $d=3$ and going towards $d=2$, it is found that intermittency is greatly reduced as soon as $d$ is less than three.

\subsection{The large $d$ limit}

These dramatic differences in dynamical behaviour between two- and three-dimensional turbulence led to further investigations into higher dimensions. These studies were typically motivated by the anomalous exponents observed in three-dimensional experiment, with the hope that the problem may simplify in the limit of large dimension. To this end, \citet{fournier1978infinite} considered the infinite dimensional limit. No obvious simplifications were found; however, they were unable to rule out the vanishing of intermittency in infinite dimensions as was found by Kraichnan for a passive scalar advected by a random velocity field. The differences between turbulence in three dimensions and in higher dimensions is far more subtle than between two and three dimensions. This is related to a lack of additional positive-definite ideal invariants for higher dimensions.

Additionally, \citet{fournier1978infinite} considered the role of pressure and the incompressibility condition in infinite dimensions. It had been suggested that the incompressibility condition should weaken as the dimension increases, leading to the possibility of turbulence tending towards Burgers equation statistics in this limit. However, in \citep{fournier1978infinite} the pressure is found to continue to play an important role even in infinite dimensions. This result has been questioned by \citet{falkovich2010new}, who found that in the infinite dimensional limit incompressible turbulence may have Burgers scaling, with the discrepancy in findings attributed to the role of Gaussian initial conditions in \citep{fournier1978infinite}. Finally, recent work by Rozali studied relativistic turbulence in a large number of spatial dimensions, where it is found that equations of motions are simplified in the limit $d\rightarrow\infty$ \citep{rozali2018holographic}.

\subsection{Numerical studies beyond three dimensions}

Given the analytical challenges turbulence poses, the use of numerical methods, in particular DNS, has been invaluable. However, only recent has it become possible to perform DNS of turbulence in greater than three dimensions. These studies began with the work of \citep{suzuki2005energy,gotoh2007statistical,yamamoto2012local} looking at DNS of decaying turbulence in four and five spatial dimensions. Due to the extreme computational cost of such simulations, the highest resolution achieved in four dimensions was $256^4$ collocation points. Before discussing the main results of these works, we wish to mention the need for caution in the interpretation of results from studies of free decay at modest resolution. In \citet{qian1997inertial} it was suggested that a true inertial range will not be observed until the Taylor Reynolds number, $\Rey_{\lambda} \geq 2000$. This is an important point because even in the highest resolution DNS of three-dimensional turbulence \citep{iyer2020scaling} performed to date, with $16,384^3$ collocation points, this value has not been achieved. Indeed, in \cite{ishihara2020second} it was observed in DNS at a Taylor Reynolds number of 2250 that the scaling range found was not viscosity independent, and thus could not be a true inertial range. Furthermore, it can be seen \citep{bos2012reynolds} that free decay of turbulence requires even higher values for an appreciable inertial range. As such, there will have been no true inertial range in these higher dimensional DNS studies making any definitive conclusion about K41 impossible at this point.  

In any case, in \citet{gotoh2007statistical} the intermittency of the dissipation rate is studied in several ways. Most notably, they looked at the total dissipation and a surrogate dissipation often used in experiment. They found reduced intermittency for the total dissipation alongside increased intermittency of the surrogate. However, the reduction of intermittency in the total dissipation was a larger effect than the increase in the surrogate. However, as discussed in the introduction, it is in fact the fluctuations of the energy transfer rate that are relevant for the validity of K41, and this quantity was not measured in \citet{gotoh2007statistical}. Measurements were also made of the longitudinal structure functions in four dimensions, where it was found that deviations were larger when compared with three dimensions. However, as the authors note, there is no inertial range present in their simulations, hence this finding cannot be considered conclusive. Indeed, it is also conceivable that the finite Reynolds number effect in four dimensions differs in strength from that in three dimensions further complicating the interpretation of the data found in \citet{gotoh2007statistical}.

In a more recent study \citet{berera2020homogeneous} performed DNS of stationary four-dimensional turbulence at an increased resolution of up to $512^4$ collocation points. Of course, even at this level of resolution no inertial range can be found, so no attempt to measure anomalous scaling was made in this work. Instead, this work focused on the role of fluctuations by considering the variance of the total energy at stationary state alongside measurements of the velocity derivative skewness and dimensionless dissipation rate. When considering these fluctuations there is a dramatic reduction when going from three to four dimensions of the order of an order of magnitude at $\Rey_L \approx 1000$, suggesting a possible simplification of the dynamics. The increase from three to four dimensions is not enough for arguments based on the central limit theorem to explain the reduction, although of course this will become a factor at higher dimensions. Once more, at the resolution available caution should be taken in the interpretation of this result, however, how these fluctuations behave with increasing dimension may be of interest considering the analogies between turbulence and critical phenomena.

At present, computing power is such that DNS of turbulence in dimensions greater than four at anything beyond low resolution is out of reach. However, through the use of closure approximations higher dimensional turbulence can be studied numerically. In general, these closure-based methods do not account for intermittency and thus are not suited for studies in anomalous scaling. However, they are well suited for the study of low order statistics in higher dimensions as was carried out in \citet{clark2021effect} using the EDQNM closure \citep{orszag1970analytical}. Here it was found that the enstrophy production reaches a maximum near six dimensions. Without DNS results it cannot be determined if this is specific to the EDQNM closure or a feature of the Navier-Stokes equations themselves, however, it suggests a possible change in dynamical behaviour that should be investigated further.

\section{Chaos and predictability in turbulence}\label{se:chaosInt}

In 1963 Lorenz found that the evolution of a simplified model of the atmosphere was highly sensitive to initial conditions \citep{lorenz1963deterministic}. Systems exhibiting this property are said to be deterministically chaotic. The study of chaotic systems and their predictability has been an area of growing interest ever since. In particular, it is well-known that turbulent phenomena are chaotic \citep{ruelle1971nature,deissler1986navier}, and given the ubiquity of turbulent flow in nature, chaos and predictability in turbulence have been widely studied. The first series of works on predictability were carried out by Leith and Kraichnan in the early 1970's. These authors used statistical closures such as the Test Field Model (TFM) and EDQNM to study this problem \citep{leith1971atmospheric,leith1972predictability}. In the past few decades, the rapid increase in computational power has allowed DNS to be used in the study of the sensitivity of turbulence to initial conditions \citep{Berera2018,Boffetta2017, mukherjee2016predictability,mohan2017scaling,ho2019chaotic, ho2020fluctuations,boffetta1997predictability,nastac2017lyapunov,li2020superfast,yoshimatsu2019error,boffetta2001predictability,clark2020chaos,clark2020chaotic}. 

Given that turbulence is chaotic, an infinitesimal perturbation $\lvert \delta u(0)\rvert$ made in an evolving field $\bm{u}(\bm{x},t)$ will grow exponentially in time such that 
\begin{equation}
\lvert\delta\bm{u}(t)\rvert \sim \exp\left(\lambda t\right) \quad ,
\label{eq:lyapunov_exponent_def}    
\end{equation}
where the growth rate $\lambda$ is the maximal Lyapunov exponent. Applying an infinitesimal perturbation in this way mimics the effect of lack of predictability due to experimental error. In 1979, using dimensional arguments Ruelle stated that for three dimensional turbulence, the maximal Lyapunov exponent (which has units of inverse time), is associated with the Kolmogorov microscale time $\tau = \left(\nu/\varepsilon\right)^{1/2}$, where $\varepsilon$ is the kinetic energy dissipation. This is the smallest time scale in the system, then $\lambda \sim \tau^{-1}$ \citep{ruelle1979microscopic}. Using K41 arguments, it can be further shown that the Lyapunov exponent is related to the Reynolds number of the flow, defined as Re$=UL/\nu$, where $L$ is the integral length scale ($L= \int dk E(k)/k$), $E(k)$ is the kinetic energy spectrum, and U is the rms velocity of the flow. This gives the following scaling relation

\begin{equation}
    \lambda \sim  \frac{1}{T} \Rey^{\alpha}\quad ,
    \label{eq:ruelle_relation}
\end{equation}
where $\alpha = 0.5$ and $T$ is the large eddy turnover time, defined as $T = L/U$. This prediction has been tested numerically using DNS \citep{Berera2018,Boffetta2017,ho2020fluctuations}.

In this work, we study the error growth for $d$-dimensions (with $d=2,3,4,5,6,7,8$) using an EDQNM closure. For the case of $d=3$ and $4$, this allows us to explore the parameter space in far greater detail than is possible for current DNS. For $d>4$, DNS is not achievable in practice, a lattice box of $16384^3$ collocation points, which is the largest one achieved by DNS to this date \citep{iyer2019circulation} is approximately equivalent in computational power to having a box with a size of $338^5$ points, which will not allow for the required scale separation for an inertial range to exist. Running DNS simulations at high $\Rey$ values come with massive computational expense, and the resolution required increases as $\Rey^{3d/4}$. As such, the problem is further exacerbated by dimension. In EDQNM, the increase in computational cost with dimensions and $\Rey$ is much slower. In this work, we explore ranges up to $\Rey = 10^5$ in an eight-dimensional space. It is worth mentioning again that intermittency corrections are not captured in the EDQNM closure model. Thus, any scaling anomalies seen in this work cannot be associated to intermittency corrections. We also note here that results from the EDQNM equation for second and third order statistical quantities are in good agreement with DNS measurements in three and four dimensions, as shown in \citep{clark2021effect}. As such, our work will focus on such quantities in determining the chaotic properties of higher dimensional flows. Other possible sources for deviations from Ruelle's scaling can be due to finite Reynolds number effects, systematic errors or the existence of a physical mechanism in which chaos is led by a characteristic timescale different from the Kolmogorov  microscale time.

\subsection{Error growth in the EDQNM model}
\label{se:error_growth_in_edqnm}

As previously mentioned, the range of $\Rey$ and $d$ that can be studied using DNS is severely limited, hence in this present work we use EDQNM closure to study error growth in $d$-dimensions. Another advantage of using this closure model is that the extension to non-integer dimensions is trivial. This can be used to analyse the features of the transition between dimensions, as was done in previous works when studying the transition between two and three dimensions, as mentioned in section \ref{se:intro}. 

We use the spectral representation of the velocity field $u_i(\bm{k},t)$, where we use index notation with $i = 1,2,\dots,d$ and summation is implied unless otherwise stated. Our system consists of two velocity fields that are initially close together $\bm{u}^{(1)}(\bm{k},t)$ and $\bm{u}^{(2)}(\bm{k},t)$, and that are statistically identical, homogeneous and isotropic. The error field is $\delta\bm{u} = \bm{u}^{(1)} - \bm{u}^{(2)}$. Then, we define the following quantities
\begin{subequations}
\begin{align}
    E(t) &\equiv  \frac{1}{2} \int d^dk \, \langle u_{\alpha}^{(n)}(\bm{k},t)\,u_{\alpha}^{(n)}(-\bm{k},t)\rangle    \quad  . \label{eq:E_def} \\
    E_W(t) &\equiv \frac{1}{2}\int d^dk \,\langle u_{\alpha}^{(n)}(\bm{k})\,u_{\alpha}^{(m)}(\bm{-k}) \rangle \quad , \label{eq:Ew_def} \\
    E_{\Delta}(t) &\equiv \frac{1}{2} \int d^dk \, \frac{1}{2}\langle \delta u_{\alpha}(\bm{k}) \,\delta u_{\alpha}(\bm{-k}) \rangle \quad  =E(t) - E_W(t)   
    \label{eq:Ed_def} \quad ,
\end{align}
\end{subequations}
where $\langle \cdot \rangle$ denotes the ensemble average, $i,j=1,2, \dots ,d$ and $m,n = 1,2$ but $m\neq n$. We denote $E$ the kinetic energy as is usual, and we will refer to $E_W$ and $E_{\Delta}$ as the correlated and decorrelated energy respectively. 


We note that due to isotropy, the quantities between angled brackets depend only on $k$ and $\int d^dk = \int dk k^{d-1}S_d$, where $S_d \equiv 2\pi^{d/2}k^{d\shortminus 1}/\Gamma\left(d/2\right)$, where $\Gamma(x)$ is the gamma function. Hence, we can define the following spectra

\begin{subequations}
\label{eq:E_spetra}
\begin{align}
    E(k,t) &= k^{d-1} \,S_d \,  \, \langle u_{\alpha}^{(n)}(\bm{k},t)\,u_{\alpha}^{(n)}(-\bm{k},t)\rangle \quad ,\label{eq:E->U} \\
    E_W(k,t) &= k^{d-1} \,S_d \,  \, \langle u_{\alpha}^{(n)}(\bm{k},t)\,u_{\alpha}^{(m)}(-\bm{k},t)\rangle\quad (m\neq n) \quad , \label{eq:E_W->W} \\
    E_{\Delta}(k,t) &= E(k,t) - E_W(k,t) \quad .\label{eq:E_delta->delta}
\end{align} 
\end{subequations}

The equations for the evolution of these energies can be derived from Navier-Stokes equations, obtaining 

\begin{subequations}
\label{eq:E_and_Ew}
\begin{align}
\begin{split}
\left[\partial_t + 2\nu k^2 \right]  E(k) &=
T(k) + f(k)
\end{split}
    \quad , \label{eq:E_equation} \\
\begin{split}
\left[\partial_t + 2\nu k^2 \right] E_W(k) & =  T_W(k) - T_X(k) +f(k)
\end{split}
  \quad , \label{eq:Ew_equation} \\
\begin{split}
\left[\partial_t + 2\nu k^2 \right] E_{\Delta}(k) & =  T_{\Delta}(k) + T_X(k) 
\end{split}
\quad , \label{eq:Edelta_equation}  
\end{align}
\end{subequations}

in which $T_{\Delta}(k) = T(k) - T_{W}(k)$, and

\begin{subequations}
\label{eq:Transfers}
\begin{align}
T(k) &=8 K_d \iint\limits_{\Omega(k)} dp dq \, \frac{k}{pq} \left(\frac{\sin\alpha_k}{k}\right)^{d\shortminus3} \theta_{kpq} \, b^{(d)}_{kpq}   \left[E(p)E(q)k^{d\shortminus1} - E(q)E(k)p^{d\shortminus1}\right] , \label{eq:T_equation} \\
T_W(k) &=  8 K_d \iint\limits_{\Omega(k)} dp dq  \,  \frac{k}{pq} \left(\frac{\sin\alpha_k}{k}\right)^{d\shortminus3}  \!\!\!\!\!\theta_{kpq} \,b^{(d)}_{kpq}    \bigl[E_W(p)E(q)k^{d\shortminus1} - E(q)E_W(k)p^{d\shortminus1} \bigr] , \label{eq:Tw_equation}  \\
T_{\Delta}(k) &=  8 K_d \iint\limits_{\Omega(k)} dp dq  \,  \frac{k}{pq} \left(\frac{\sin\alpha_k}{k}\right)^{d\shortminus3}  \!\!\!\!\!\theta_{kpq} \,b^{(d)}_{kpq}   \bigl[E_{\Delta}(p)E(q)k^{d\shortminus1} - E(q)E_{\Delta}(k)p^{d\shortminus1} \bigr] , \label{eq:Tdelta_equation}  \\
T_X(k) &=  8 K_d \iint\limits_{\Omega(k)} dp dq  \,  \frac{k}{pq} \left(\frac{\sin\alpha_k}{k}\right)^{d\shortminus3}  \theta_{kpq} \,b^{(d)}_{kpq}  E_W(p)E_{\Delta}(q)k^{d\shortminus1} , \label{eq:Tx_equation}  
\end{align}
\end{subequations}
where $K_d =  S_{d-1} / S_d(d-1)^2$ and $\alpha_k$ is the opposite to $\bm{k}$ in the triad formed by $\bm{k}+\bm{p}+\bm{q} = 0$ and hence, $\sin^2\alpha_k = 1- \left(p^2+q^2-k^2\right)^2/4p^2q^2$. The term $f(k)$ is a forcing function, that must be the same in both (\ref{eq:E_equation}) and (\ref{eq:Ew_equation}), otherwise error growth will be also driven by the forcing difference. The factor $\theta_{kpq}$ is related to the relaxation time of the triad, which is computed using the eddy-damped Markovian approximation
\begin{align}
    \theta_{kpq} &= \frac{1-\exp\left(-(\mu_k+\mu_p+\mu_q)\,t\right)}{\mu_k + \mu_p +\mu_q} \quad ,\\
    \mu_r = &= \nu k^2 + \beta \sqrt{\int_0^k\, ds \,s^2E(s)} \quad,
    \label{eq:theta_definition}
\end{align}
where $\beta$ is a free parameter of the model that is adjusted for each dimension to obtain the correct Kolmogorov constant, $C_d$ in the energy spectrum which is of the form shown in (\ref{eq: k41}). After using such an approximation, this form for $\theta_{kpq}$ has a number of properties that make it suitable for numerical simulations as discussed in \citep{leith1971atmospheric}. We note here that as in previous studies we make the arbitrary choice to use the same eddy damping factor for both $E(k)$ and $E_W(k)$ equations.
The factors $b^{(d)}_{kpq}$ are related to the triad geometry. 

\begin{align}
b^{(d)}_{kpq} &= \frac{p}{2k} \left[(d-3)Z + (d-1)XY + 2Z^3\right] \quad, \label{eq:b_definition}\\
X &= -\frac{\bm{p}\cdot\bm{q}}{pq} \qquad Y = -\frac{\bm{q}\cdot\bm{k}}{qk} \qquad Z = -\frac{\bm{k}\cdot\bm{p}}{kp} \quad ,\label{eq:xyz_definitions}
\end{align}

It is important to note that equations (\ref{eq:E_and_Ew}) are not chaotic themselves. These simply describe the statistical closure of Navier-Stokes equations and the error growth produced by two infinitesimally close initial conditions. Since Navier-Stokes are known to be chaotic, we expect our EDQNM closure equations to reproduce their main statistical aspects and the error growth associated.

One last important detail for the EDQNM approximation is that the values used for the free parameter $\beta$ for arbitrary dimension $d$ are obtained from previous numerical works. The values of the Kolmogorov constant $C_3$ and $C_4$ in three and four dimensions, respectively, are obtained from DNS simulations in \citep{berera2020homogeneous}. For higher dimensions, there is no DNS data available, instead values the Kolmogorov constants $C_d$ are obtained from simulations using the Lagrangian renormalised approximation (LRA)  in \citep{gotoh2007statistical}, that is another closure approximation that does not depend on the choice of any parameter. See the appendix of \citep{clark2021effect} for information of how the free parameter should be set to obtain a given value for the Kolmogorov constant in each dimension. Additionally, since the choice of free parameter is based on weaker evidence especially for higher dimensions, we performed regular consistency checks, especially to ensure that the main result of this work, i.e. the existence of a critical dimension $d_c$ at which the system transitions from chaotic to non-chaotic behaviour, does not depend on the choice of the free parameter $\beta$. For $d=6$ the value of the free parameter obtained by the criteria mentioned above is $\beta = 0.23$, and if we vary for instance this value to $\beta \approx 0.16-0.3$, the Kolmogorov constant changes less than 10\% and the system present non-chaotic behaviour at all times. However, for lower values, for instance $\beta = 0.11$ and below, the system changes its qualitative behaviour and the system presents now non-chaotic behaviour. Nevertheless, we performed the same test in $d=7$ and we observed that regardless of the choice of free parameter, the system is always non-chaotic for this dimension. This indicates that the choice of free parameter in this paper will only affect the dimension $d_c$ at which the systems transition from chaotic to non-chaotic behaviour, but not the general picture.

The numerical procedure consists in computing the integrals in Eqs. (\ref{eq:E_and_Ew}) following the scheme outlined by \citet{leith1971atmospheric} and then using a predictor-correlator time stepping method to evolve $E$ and $E_W$ in time. The wavenumber space is discretised using $N$ real numbers whose logarithm is evenly spaced, then $k_i = 2^{i/F}$ for $i = 0,\dots,N-1$. The time and spatial resolution must be increased for growing $\Rey$ and $d$ to obtain stable results. More details of the numerical method can be found in \citep{clark2021effect}.

The factor $\sin \alpha_k$ in Eq. ($\ref{eq:Transfers}$) adds a further complication since its value varies abruptly within $\Omega(k)$ and it is not defined outside this region. To deal with the first issue, the value of $F$ is set high enough to ensure that the numerical results are stable. An additional complication comes from points which lie outside the integration domain, but some part of their associated volume lies within the domain. For such points $\sin\alpha_k$ is not defined, hence  these points are removed from the summation. Any issues arising from removing these points can be mitigated by choosing a large value for $F$.    

The resolution of the numerical integration scheme is such that the correct Kolmogorov constant at each dimension is achieved, this criteria is also used in \citep{clark2021effect}. Still, this condition is not enough for equations (\ref{eq:E_W->W}) and (\ref{eq:E_delta->delta}), since larger temporal and spatial resolution are needed for the net transfer $T_X$. The temporal resolution needs to be increased as $\Rey$ grows due to the rapid temporal variations in $E_{W}(k,t)$ as the system evolves, which are of the order of the microscale time $\tau$ following Ruelle's relation. Hence, we increased the resolution by varying $F$ and the timestep $dt$ until stable results were obtained for the growth or decay rate. For moderate values of $\Rey$, we cross-checked our log spaced grid runs against the same run with linearly spaced grid to ensure the simulation is well-resolved. For larger Reynolds number, the simulations become computationally too expensive to perform this procedure.

The form of the forcing for every run is 
\begin{equation}
    f(k) = \begin{cases}
        \dfrac{\varepsilon}{\Delta_k} \qquad &\text{if} \qquad k \leq \Delta_k \\
        0 \qquad &\text{if}  \qquad k > \Delta_k \quad, 
    \end{cases}
\end{equation}
where $\Delta_k$ is the forcing band, set to $\Delta_k = 2$ in all runs. This forcing has the advantage of allowing us to set the dissipation rate $\varepsilon$ \textit{a priori}. Throughout this work we set $\varepsilon=0.1$. 

Each run consists in evolving  $E(k,t)$ until it reaches the steady state, then, the energy correlation $E_W(k,t)$ is initialised by introducing a perturbation so that the initial spectrum of the error is a function of the form 
\begin{equation}
    E_{\Delta}(k,t=0) = \frac{A}{1+\exp\left(-\frac{4(k-k_p)}{k_p}\right)} \quad,
\end{equation} 
which makes the perturbation stronger for the large wavenumbers over the low ones. The constant $A$ is set for each simulations such that $E_{\Delta}(t=0) = 10^{-7}$. The value of $k_p$ is set at approximately $k_p \approx 0.9k_{\text{max}}$ in all runs, where $k_{\text{max}}$ is the maximum wavenumber simulated, chosen such that $k_{\max}\eta >2$, where $\eta$ is the Kolmogorov length scale $\eta=\left( \nu^3/\varepsilon \right)^{1/4}$. The choice of $k_p$ makes no significant difference in the evolution of $E_{\Delta}(k,t)$. The maximal Lyapunov exponent is taken from the evolution of $E_{\Delta}(t)$ in our simulations. We note that from its definition (\ref{eq:Ed_def}), we have $|\delta\bm{u}(t)| = \left(2E_{\Delta}(t)\right)^{1/2}$ and thus we have 
\begin{equation}
E_{\Delta}(t) \sim \exp(2\lambda t) \quad .
\end{equation}
Even though the Lyapunov exponent is defined for infinitesimal perturbations, in numerical works it is common practice to use small perturbations that tend to the Lyapunov exponent as the perturbation becomes smaller \citep{Boffetta2017}. For instance, the finite time Lyapunov exponent (FTLE) is defined by the measure of error growth in a fixed time interval, whereas the finite size Lyapunov exponent is defined similarly by fixing the growth magnitude (eg. double the magnitude of the initial perturbation) and then measuring the time it takes to reach that from an initial small perturbation. In both cases, the Lyapunov exponent is defined as $\left(2 \Delta t \right)^{-1}\log\left(E_{\Delta}(\Delta t)/E_{\Delta}(0)\right) $. In our EDQNM simulations, given the wide range of timescales involved in the system, there is no systematic way to fix either a magnitude or a timescale. Hence, we manually find the latest stage of exponential growth or decay, and then perform a linear fit of $(1/2)log(E_{\Delta}(t)/E))$ vs. $t$, where $E$ is the total energy defined in the steady state as $E=U^2/2$. Further discussion can be found on section \ref{se:evolution_for_d>5}.
Most of the DNS results for three dimensions are taken from previous work \citep{Berera2018,ho2020fluctuations}, except the ones with $\Rey<36$. All the four-dimensional DNS simulations are first shown in the present work. The DNS code consist in evolving the forced NSE, using a fully de-aliased pseudospectral code in a periodic lattice with $N^3$ and $N^4$ collocation points for three and four spatial dimensions respectively. In our work we use values $N = 64, \, 128$ and $256$. The same second order predictor-correlator time stepping method as in the EDQNM runs is used. In this work, we measure the Lyapunov exponents using the FTLE procedure, described above. For details on the DNS code refer to \citep{yoffe2013investigation} and \citep{rho2019}, and for details on the method to obtain FTLE refer to \citep{ho2020fluctuations}.

\section{Results}
\label{se:results}

In this section we analyse the error growth in dimensions $d=3,4,5,6,7$ and $8$ as well as between five and six dimensions. Given the low computational cost of EDQNM simulations compared to DNS, we can explore the chaotic properties in a much wider range of the parameter space. We run several simulations with values of $\Rey$ up to $5 \cdot 10^5$. This Reynolds number corresponds to the integral scale and is defined as in \citep{clark2021effect} by \begin{subequations}
\label{eq:Re_L}
\begin{align}
\Rey &= \frac{UL}{\nu}
\quad , \label{eq:Re} \\
L_{d}  &= \frac{\Gamma\left(\frac{d}{2}\right)\sqrt{\pi}}{\Gamma\left(\frac{d+1}{2}\right)u^2}\int_{0}^{\infty}\mathrm{d}k \, E(k)k^{-1}
 \quad , \label{eq:L}  \\
 U^2  &= \frac{2}{d}\int_{0}^{\infty}\mathrm{d}k \, E(k)\
\quad . \label{eq:Re1}
\end{align}
\end{subequations} This value is above the value of $\Rey$ obtained in the largest DNS for isotropic turbulence to date, that reaches a value of $\Rey \approx 2 \cdot 10^5$ in runs with short averaging times \citep{iyer2019circulation}. This $\Rey$ value is substantially lower than in \citet{clark2021effect}, due to the higher temporal resolution requirements of the $E_{W}(k,t)$ equation as $\Rey$ is increased. It is important to recall that although the EDQNM approximation gives a generally good agreement with DNS results, it is based on a closure hypothesis, thus some features such as intermittency (that is captured in DNS) are not present in the EDQNM model. Also, this model only uses moments up to third order, hence any effect present in higher order moments of the velocity field will not be captured.

Our main result is that the error in the evolution of two fields in NSE using the EDQNM closure, does not grow above a critical dimension $d_c \approx 5.88$. The error $E_{\Delta}(t)$ of solutions with $d>d_c$, undergo an initial stage of fast exponential growth, followed by an exponential decay of lower rate, that corresponds to a non-chaotic behaviour. Notably, in the integral dimensions below $d_c$ we find that the Ruelle scaling holds, however, the value for the exponents themselves drops with dimension. As we will demonstrate in section \ref{se:chaos_d_dim}, there is some discrepancy between the absolute values of the Lyapunov exponents measured in the EDQNM calculations and those from our DNS. This discrepancy is larger in three dimensions, with the four dimensional EDQNM and DNS results being closer. Importantly, the trend of a smaller value for the Lyapunov exponent at a given Reynolds number as $d$ is increased is seen in both EDQNM and DNS. This provides additional credibility to the further decreases we see as we approach $d_c$. However, given the small differences seen, it is likely that the location of $d_c$, if it exists, for the Navier-Stokes equations will be different than found in EDQNM.

\subsection{Reynolds number scaling of Lyapunov exponents}
\label{se:chaos_d_dim}

\begin{figure}
 \begin{minipage}[b]{0.45\linewidth}
    \centering
    \includegraphics[width=\textwidth]{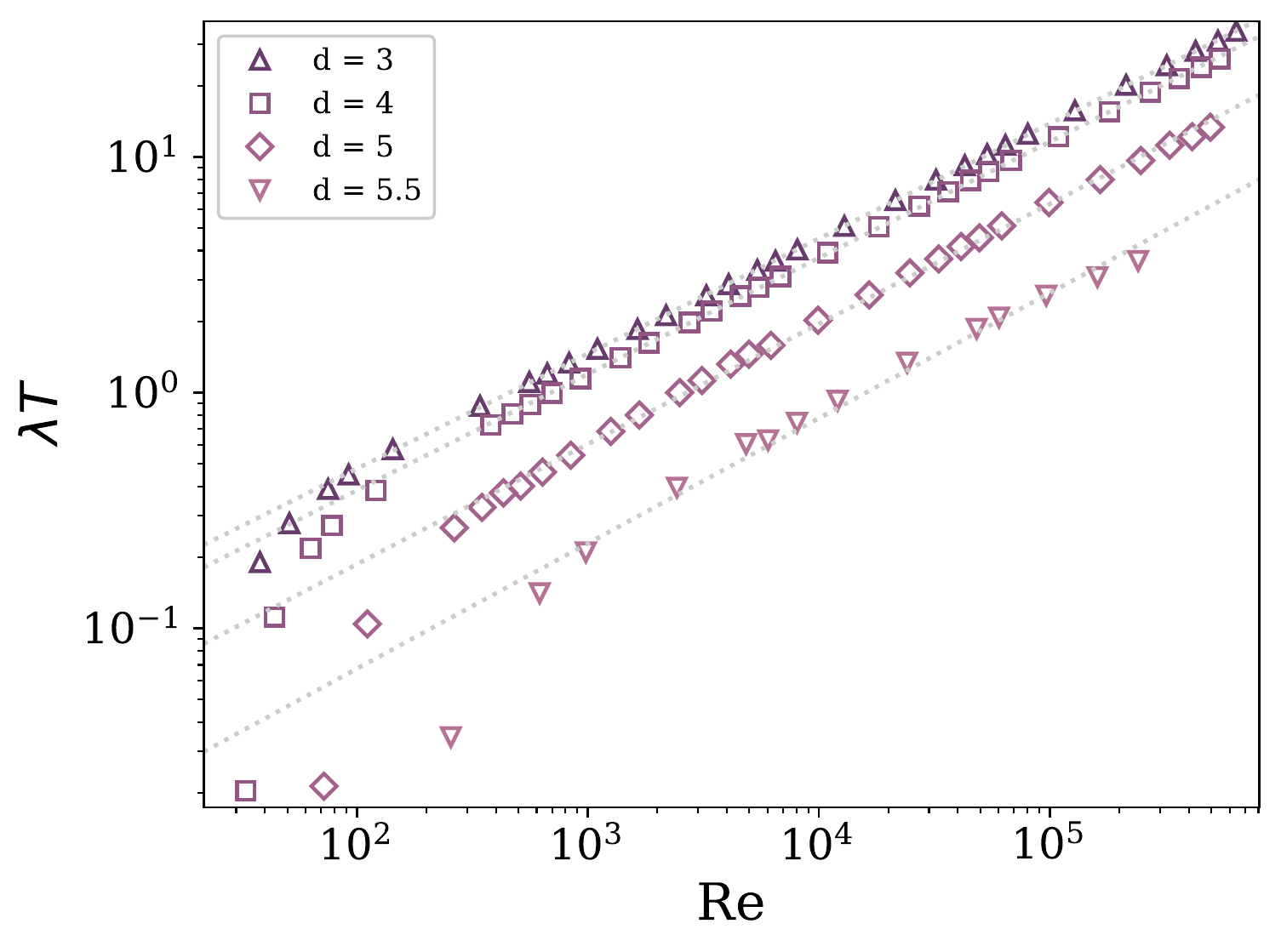}
    \caption{Scaling of Lyapunov exponents $\lambda$ with $\Rey$ for dimensions $d = 3$, $4$, $5$ and $5.5$ using the EDQNM closure.}
    \label{fig:Re_scaling_345a}
\end{minipage}
\hspace{0.5cm}
\begin{minipage}[b]{0.45\linewidth}
    \centering
    \includegraphics[width=\textwidth]{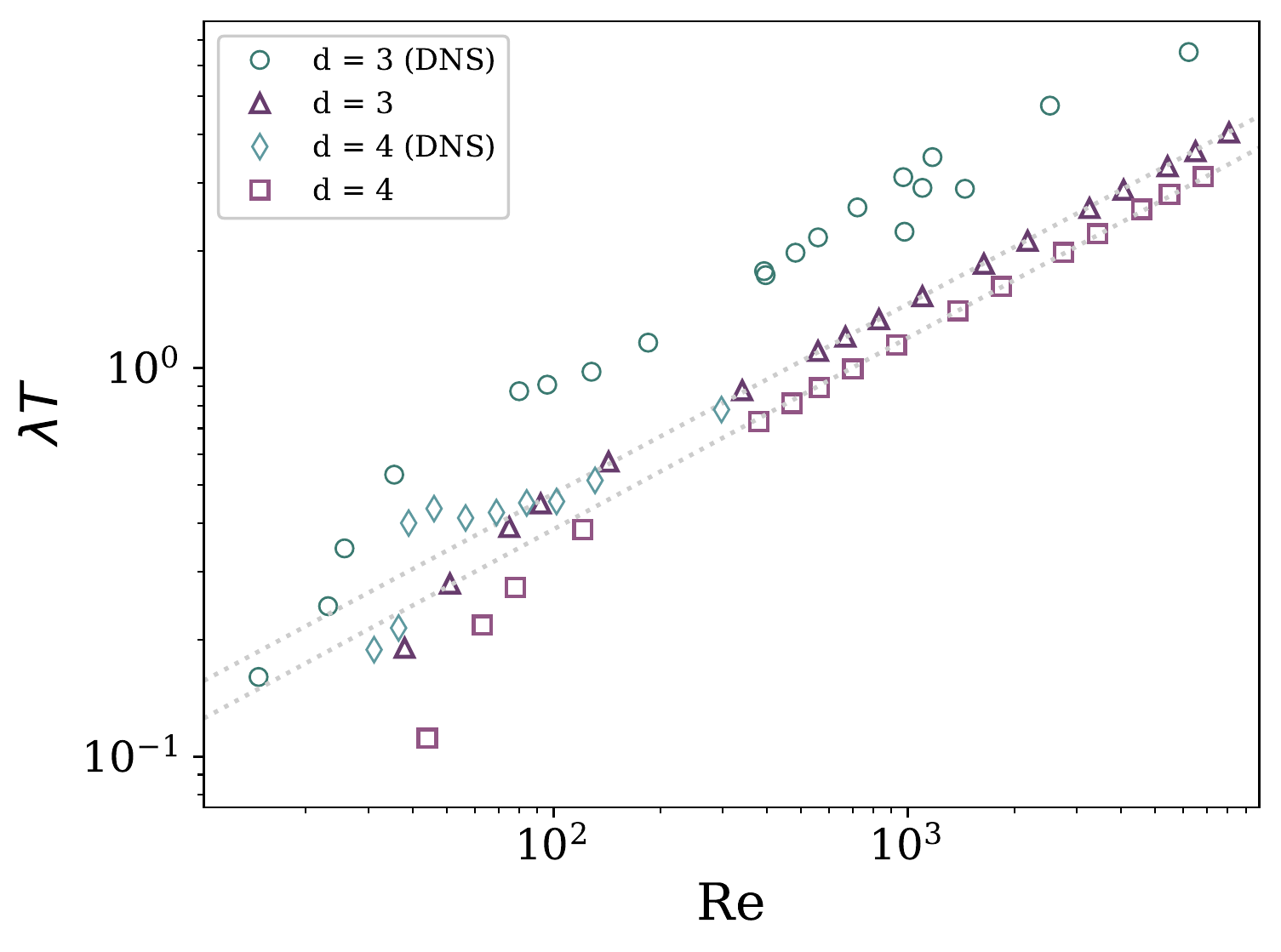}
    \caption{Scaling of Lyapunov exponents for both EDQNM and DNS in $d=3$ and $4$ in the region $\Rey < 5000$}
    \label{fig:Re_scaling_345}
\end{minipage}
\end{figure}

First, we consider the dependence of the maximal Lyapunov exponent, $\lambda$, on the Reynolds number. Figure \ref{fig:Re_scaling_345a} shows the data obtained from approximately 
thirty simulations for each dimension $d=3,4$,$5$ and $5.5$ varying the viscosity to explore a wider range of $\Rey$. For these we use the EDQNM approximation. A scaling of the form of Eq. (\ref{eq:ruelle_relation}) is found, with $\alpha_{d=3}=0.49\pm0.02$, $\alpha_{d=4}=0.50 \pm 0.02$, $\alpha_{d=5} = \alpha_{d=5.5} = 0.53 \pm 0.02$. Clearly, for all four cases Ruelle's prediction seems to hold. For $d=3$ the scaling agrees with the Ruelle dimensional prediction with the value of 0.5 being within the error bounds.

We then look at figure \ref{fig:Re_scaling_345} and we compare the results obtained with previous DNS results. We do this comparison using the EDQNM runs that have a similar range of $\Rey$ values to those used in DNS. Most of the DNS results for $d=3$ are those obtained previously in \citep{Berera2018,ho2020fluctuations}, with the additions of a few points at low Reynolds number. All the DNS results for $d=4$ are first reported in this paper. The Ruelle scaling is observed in $d=3$ with a scaling exponent $\alpha_{d=3_\text{DNS}} = 0.53 \pm 0.02$, whereas for $d=4$, there is not enough correlation in the data to perform a fit. It is likely that this is a consequence of the lack of inertial range at very low Reynolds numbers. For DNS in $d=4$, higher values of $\Rey$ would require a large computational power that is not available at the moment. This view is supported by the fact that the points seem to get closer to a Ruelle-like scaling fit after $\Rey \approx 100$.

We can make three observations from this comparison. First, the Lyapunov exponents in $d=4$ obtained using DNS are of the same order of those obtained through EDQNM, whereas for $d=3$ the DNS values of $\lambda$ are twice as big as the EDQNM ones. This indicates that although the scaling properties are well predicted by EDQNM, the actual values of Lyapunov exponents might not be accurate. The second observation is that the Lyapunov exponents measured in DNS for $d=4$ show a decrease respect to those for $d=3$. This not only supports the findings obtained using EDQNM, but also sets the question of whether DNS simulations will possess a critical dimension after which Lyapunov exponents become negative, similarly as in EDQNM as we will show in section \ref{se:evolution_for_d>5}. The importance of the DNS results for $d=4$ is that it shows that the question about the critical dimension for error growth is not only relevant in the context of the EDQNM model, but also it could be relevant for DNS and thus a property of the Navier-Stokes equations themselves. The third and last observation is that Ruelle's relation breaks down at low $\Rey$, and the qualitative behaviour in this regime is the same for all the dimensions displayed in EDQNM and in those using DNS. This is because at low $\Rey$ there is no longer a scaling range and hence no separation between the viscous and large-eddy timescales which is required for Ruelle scaling to hold.

\subsection{Error decay and critical dimension in $d > 5$}
\label{se:evolution_for_d>5}

In this section we describe the most important observation in this paper, the existence of a critical dimension $d_c$ between $5$ and $6$, above which the behaviour of the error $E_{\Delta}(t)$ stops growing and instead it decays. Note, we are not suggesting this as a critical dimension where anomalous scaling vanishes in the spirit discussed in section \ref{se:ddim}, but merely as a critical dimension for error growth. A deeper connection between the two ideas may exist, and we will discuss this possibility, but closure calculations are not suited to give a definitive answer.

\begin{figure}
    \centering
    \includegraphics[width = 0.54\textwidth]{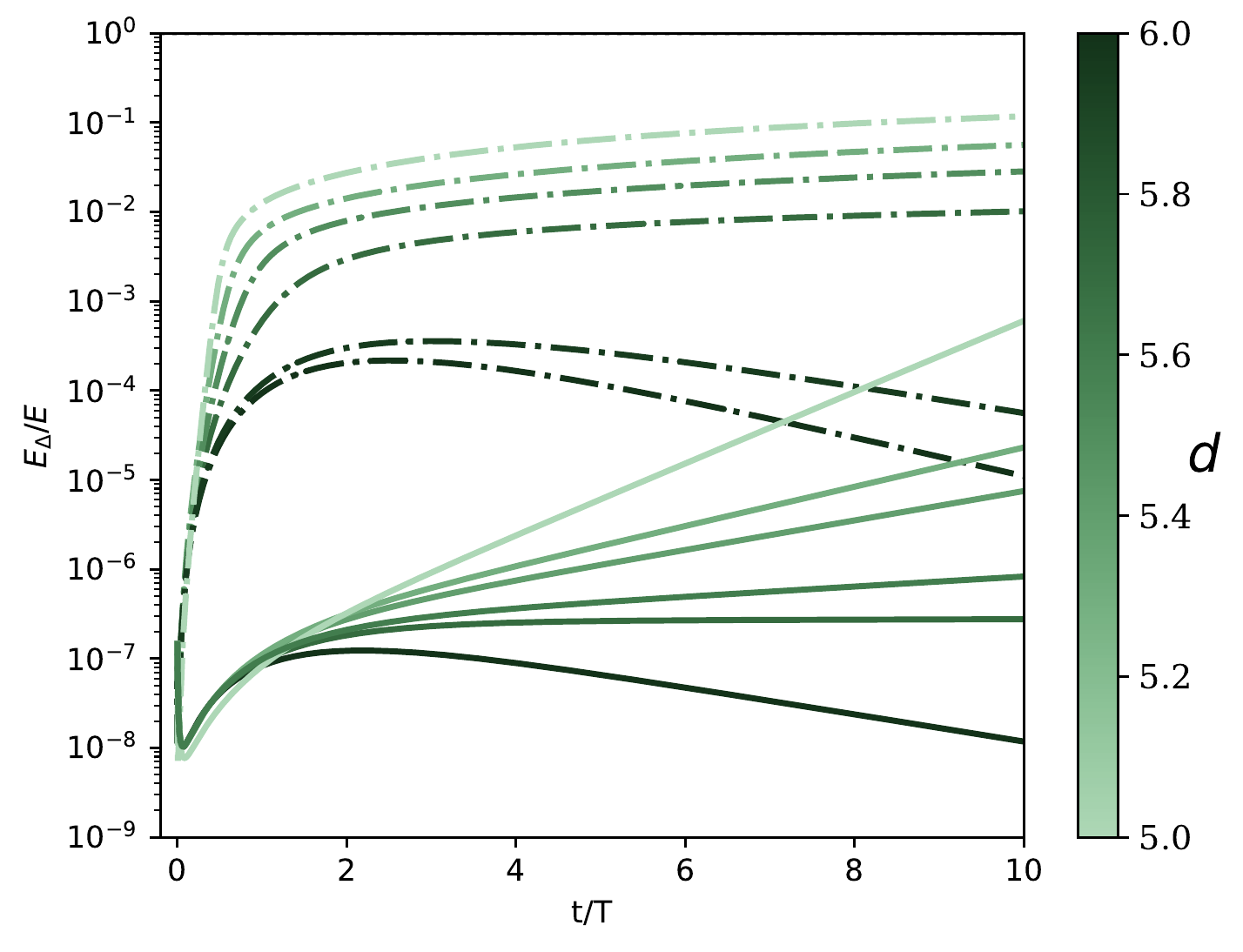}
    \caption{Fractional error growth for dimensions $5 < d < 6 $ for runs with low $\Rey$ ($\nu = 8 \cdot 10^{-4} $) represented by solid lines and high $\Rey$ ($\nu = 2 \cdot 10^{-6}$) represented by dash-dotted lines. As indicated, the line colour becomes darker moving from 5 to 6 dimensions.}
    \label{fig:e_growths}
\end{figure}

For cases above $d_c$ we observe a fast initial growth of the error, however, $E_{\Delta}(t)$ then begins to decay, at odds with what is seen below $d_c$. This initial rapid growth may be the short time super exponential error growth discussed in \citet{li2020superfast}. This transition from error growth to decay is shown in figure \ref{fig:e_growths} for a low and a high Reynolds numbers. It is seen that for lower values of $\Rey$, the fractional error growth $E_{\Delta}(t)/E$ has an initial growth stage similar for all dimensions. After this short transient, the curve reaches a stable exponential growth for values closer to $d=5$ reaching saturation at values lower than 1. When $d$ increases, this growth rate decreases gradually becoming a decay for values of $d \approx 5.71$. For larger $\Rey$, the picture is slightly different. The initial growth is fast, and it is also present for all dimensions, but it reaches higher values of $E_{\Delta}/E$. For the values of $d$ closer to 5, the growth saturates at values of $E_{\Delta}/E \approx 10^{-1} $ - $10^{-2}$. These are lower than 1 (i.e. not fully decorrelated) for reasons we will analyse in section \ref{se:error_spectrum}. As $d$ increases, the growth reaches lower values of saturation. Above dimensions $d \approx 5.8$ the growth reaches values of $E_{\Delta}/E \approx 10^{-4}$ followed by an exponential decay with a lower rate than the previous growth.

As we discussed in section \ref{se:error_growth_in_edqnm}, there are different stages with different time scales involved in the error growth. Figure \ref{fig:e_growths} shows this in more detail. We can see that for low Reynolds number and $d<d_c$, there is a rapid initial transient of growth that lasts approximately one eddy turnover time that is similar for all dimensions between 5 and 6. This stage is followed by an exponential growth that takes more than 10 large eddy turnover times to reach saturation. When the spatial dimension is increased, this last stage of exponential growth turns gradually into a decaying stage. The situation for our large $\Rey$ simulation is somehow different. The rapid initial transient of error growth is no longer present, since the timescale of the exponential growth is of the order of an eddy turnover time. This exponential growth is present in all simulations, even those with a late decaying stage. Furthermore, in these runs, it is the saturation stage that gradually turns into a decaying stage as we increase dimension. This indicates that the timescale of the error decay is approximately the same regardless of the Reynolds number, and that it becomes dominant after a few eddy turnover times. On the contrary, the exponential growth is dependent on the value of $\Rey$ following Ruelle's relation, but overcome by the decaying mechanisms at later times.

From this analysis, we note that the difference in the timescales between exponential growth and decay suggests that there are different mechanisms for both processes, and this explains why Ruelle's relation is still valid for $d=5.5$ near the critical dimension.

\begin{figure}
    \centering
    \includegraphics[width = 0.5 \textwidth]{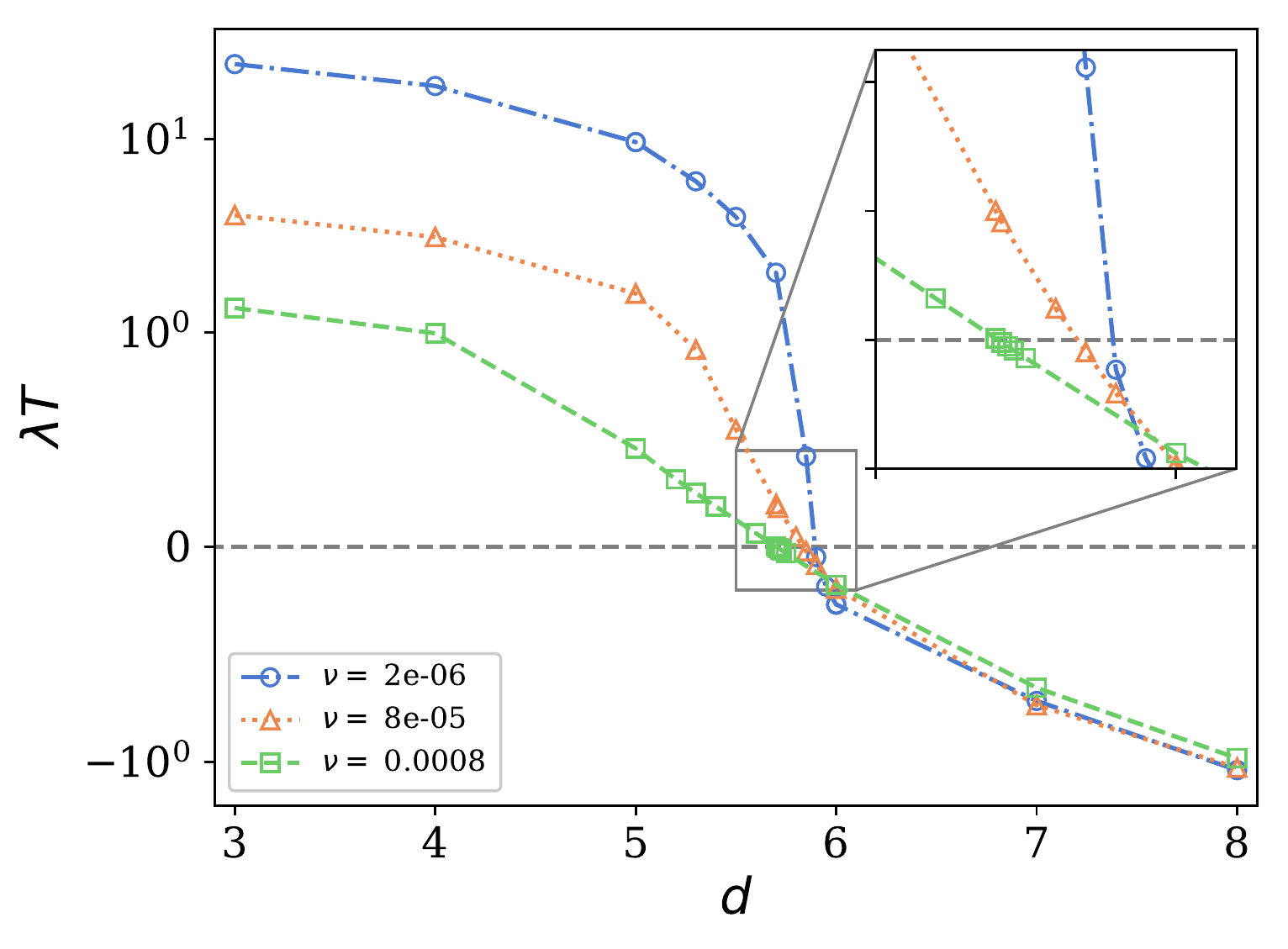}
    \caption{Lyapunov exponents $\lambda$ vs. dimension $d$ for three different viscosities. $\nu = 2 \cdot 10^{-6}$ (high $\Rey$ - circles), $\nu = 8\cdot10^{-5}$ (middle $\Rey$ - triangles) and $\nu = 8 \cdot 10^{-4}$ (low $\Rey$ - squares). Non-integer values for $d$ are shown in the transition region from 5 to 6 dimensions. The vertical axis is linear for values of $\lambda T \in [-1,1]$ and logarithmic outside that region.}
    \label{fig:lyap_vs_d}
\end{figure}

Given the $\Rey$ dependence of the transition between error growth and decay which is seen in figure \ref{fig:e_growths}, it is possible that we simply require higher and higher Reynolds numbers for error growth as the dimension is increased. To further probe this effect, we consider three different cases each with a different viscosity value. Holding these viscosities constant, we then vary the dimension as shown in figure \ref{fig:lyap_vs_d}. It is clear from this figure that as $d$ increases the value of the exponent decreases. In the zoomed in portion of the figure we can see the $\Rey$ dependence observed in figure \ref{fig:e_growths}. Interestingly, as the viscosity is lowered and the Reynolds number is raised, the drop in exponent becomes more drastic near six dimensions. This suggests the possibility of a critical value for $d$ above which error growth is not possible at finite $\Rey$.

\begin{figure}
    \centering
    \includegraphics[width = 0.54\textwidth]{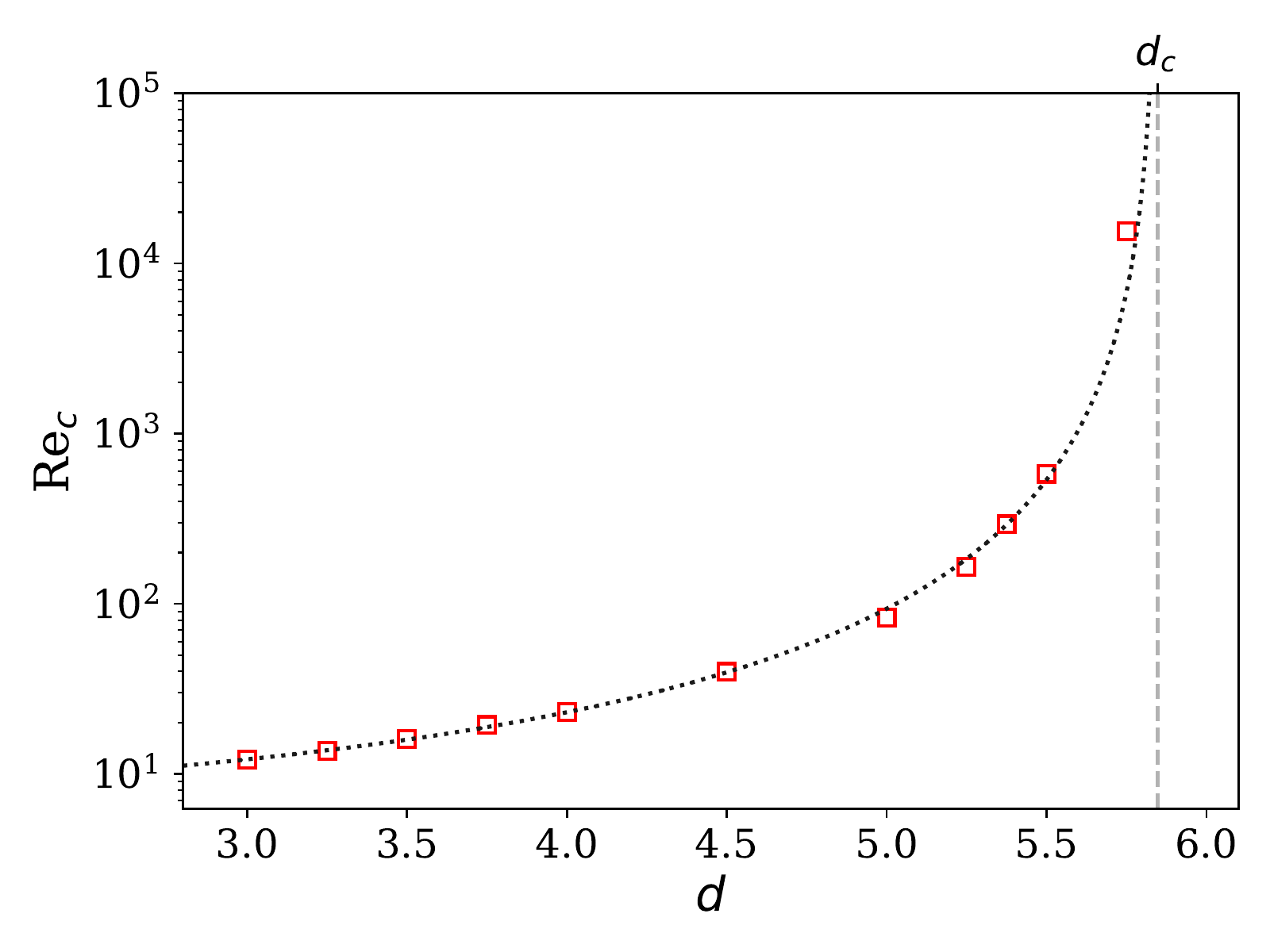}
    \caption{Critical Reynolds number vs dimension. The dotted line corresponds to the curve fit $Re_c \sim (d-d_c)^{-2}.$ }
    \label{fig:dimCrit}
\end{figure}

This behaviour raises the possibility of critical $\Rey$ values in each dimension above which error growth is possible. We thus performed a number of calculations at low $\Rey$ in three dimensions and found that for $\Rey < \Rey_c \approx 12$ the error always decays, while above this value it would always grow exponentially. Repeating this process while increasing $d$ leads to figure \ref{fig:dimCrit}. Here we find that as the spatial dimension increases this critical Reynolds number increases drastically. We find the increase is well fit by the functional form \begin{equation}
\Rey_c(d) \sim \frac{1}{\left(d-d_c\right)^2} \quad,
\end{equation} and we find $d_c \approx 5.88$. We do not know \textit{a priori} what $\Rey_{c}$ is for a given dimension, and finding it quickly becomes unfeasible due to the rapid divergence as a function of d. We will, further discuss this critical dimension for error growth in section \ref{se:error_spectrum}.

\begin{figure}
    \centering
    \includegraphics[width = 0.5\textwidth]{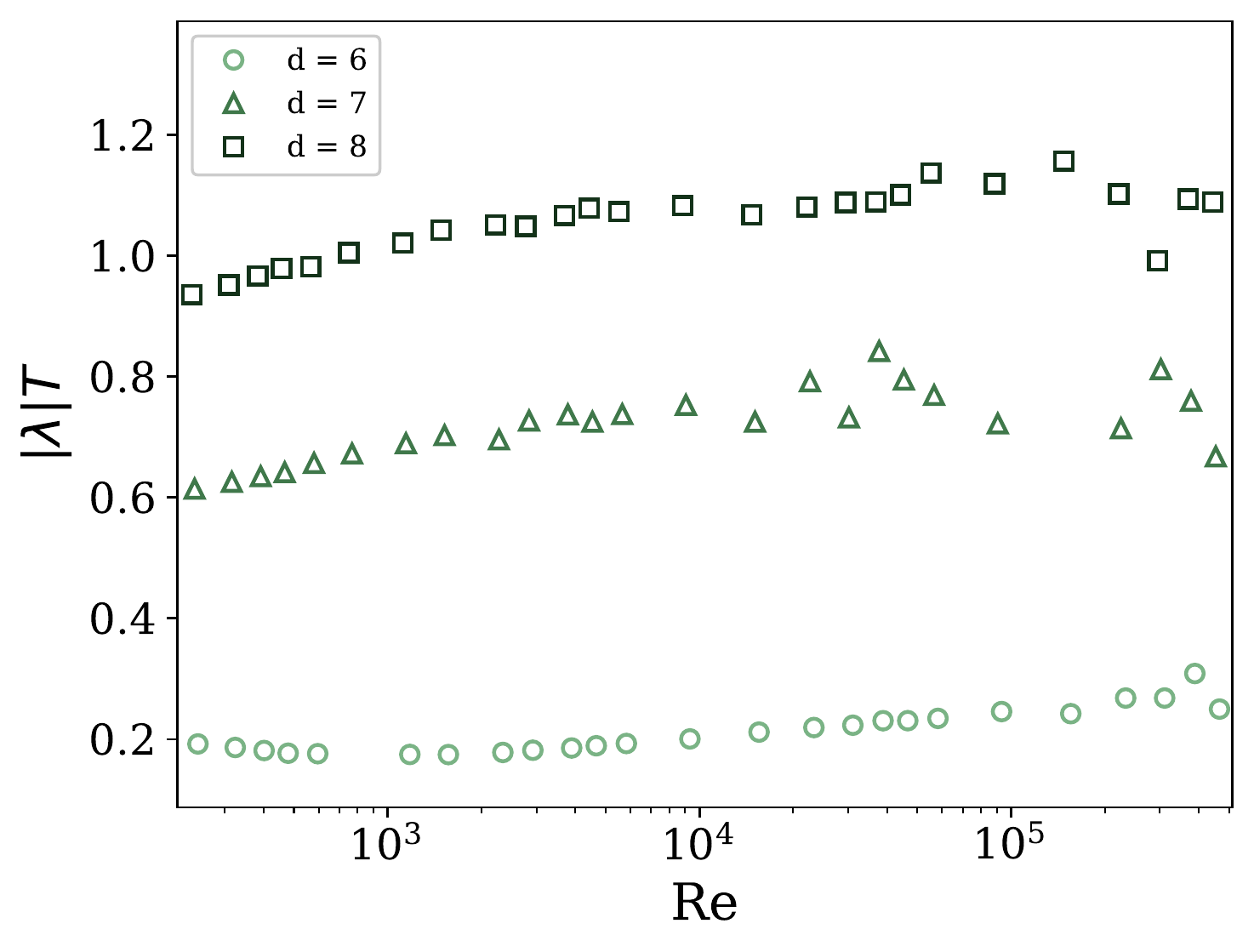}
    \caption{Absolute value of the error decay rate and its dependence on $\Rey$ for dimensions $d=6$ (circles), $7$ (triangles) and $8$ (squares).}
    \label{fig:Re_scaling_678}
\end{figure}

For completeness we show the behaviour of the Lyapunov exponents against $\Rey$ for $d = 6, 7$ and $8$ in figure \ref{fig:Re_scaling_678}. Note, we show here the absolute value of $\lambda$  and use a linear scale for the y-axis. Contrary to the case for $d<d_c$ in which error grows and the growth rate depends on the Reynolds number, the error decay rate does not have a dependence on $\Rey$ for $d>d_c$, as expected given the divergence of $\Rey_c$ shown in figure \ref{fig:dimCrit}. Physically we can understand this by considering the timescales of the flow, for example, the $\Rey$ dependence in the error growth rate for $d<d_c$ is related to the fact that the maximum rate is associated to the smallest timescales $\tau$, which has $\Rey$ dependence. In the cases of error decays, it is not clear which timescale is appropriate, however it appears this timescale must in contrast be $\Rey$ independent.

\subsection{Error spectrum}
\label{se:error_spectrum}

We can further investigate this transition to error decay by analysing the spectrum of $E_{\Delta}$. The properties of the spectrum $E_{\Delta}(k,t)$ have been studied in the past \citep{leith1971atmospheric,leith1972predictability, Boffetta2017, Berera2018}. In these works, it was observed that during the exponential growth stage, the error spectrum adopts a particular shape that peaks in the dissipation range and grows in a self-similar way. Importantly the decorrelation between the two solutions first happens at wavenumbers in the dissipation range, before then reaching smaller wavenumbers as time evolves. Finally, a K41 spectrum is obtained when the two fields are fully decorrelated. Another advantage of studying the transition from this perspective is to understand why the ratio $r(t) = E_{\Delta}(t)/E(t)$ saturates at values lower than $1$, and why this saturation value decreases with increasing $d$, as seen in the curves of figure \ref{fig:e_growths}.

\begin{figure}
\begin{minipage}[b]{0.45\linewidth}
\centering
\includegraphics[width=\textwidth]{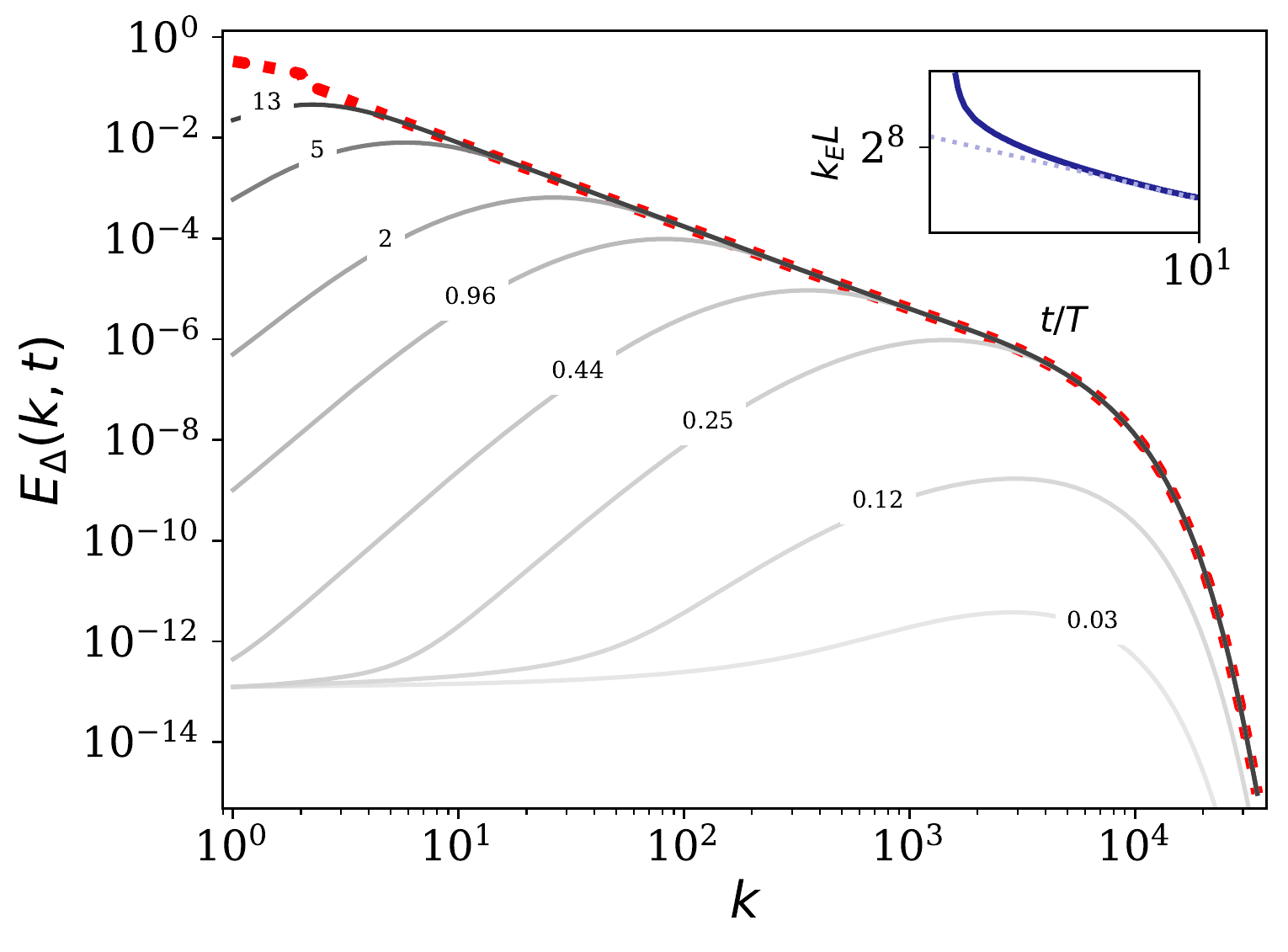}
\caption{Evolution of $E_{\Delta}(k,t)$ spectrum for a run with $d=3$ and $\nu = 1.2\cdot 10^{-6}$, where the labelled solid gray lines correspond to different large eddy turnover times $t/T = 0.03$, $0.12$, $0.25$, $0.44$, $0.96$, $2$, $5$ and $13$. The steady state spectrum $E(k)$ is represented by the red (dotted) line. The inset shows the evolution of $k_E(t)$, and the dotted line here displays the $-3/2$ scaling.}
\label{fig:spec_evolution}
\end{minipage}
\hspace{0.5cm}
\begin{minipage}[b]{0.45\linewidth}
\centering
\includegraphics[width=\textwidth]{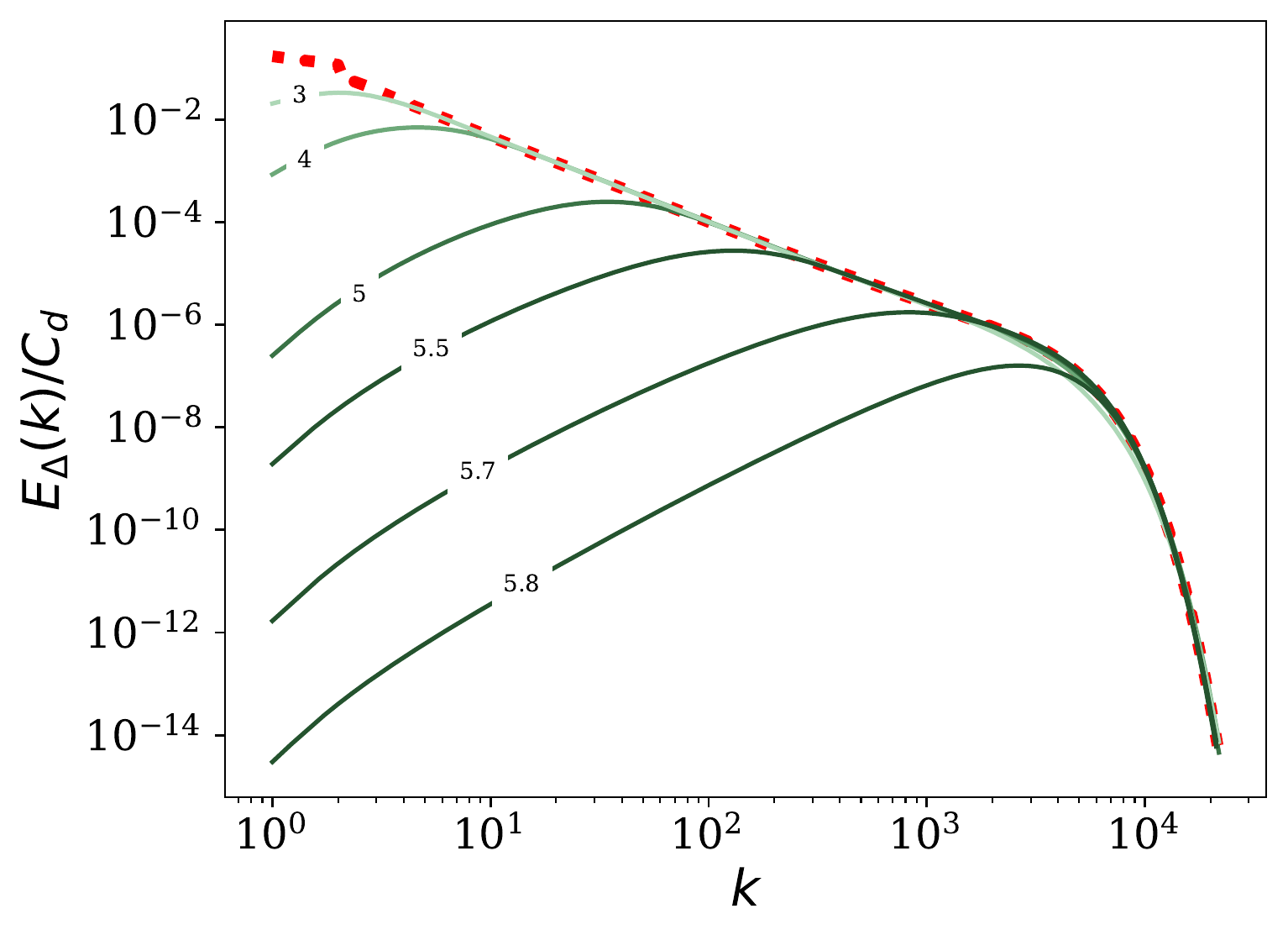}
\caption{Final spectrum of $E_{\Delta}(k)$ where the labelled (solid) lines correspond to different dimensions with $d = 3$, $4$, $5$, $5.5$, $5.7$ and $5.8$, together with the steady state spectrum of $E(k)$ represented by the red (dotted) line. All runs have the same viscosity $\nu = 2\cdot 10^{-6}$. This viscosity value corresponds to the largest Reynolds number explored in each dimension.}
\label{fig:spec_vs_d}
\end{minipage}
\end{figure}

In figure \ref{fig:spec_evolution}, we see the evolution in time of the error spectrum for $d=3$. We note that there is an initial growth that peaks just in the transition between inertial and dissipative ranges, and at this time the error grows exponentially. The large wavenumbers are decorrelated first, then as time increases lower wavenumbers become decorrelated moving through the inertial range, until finally, the error spectrum becomes approximately equal to the energy spectrum $E_{\Delta}(k,t) \approx E(k,t)$. At this point we must have the correlation energy practically vanishing i.e. $E_W(k,t)\approx 0$. This picture is known from previous works using DNS \citep{Boffetta2017,Berera2018}. Also, we observe that the data agrees with a prediction made by Lorenz which states that the time it takes for the decorrelation to reach a certain length scale $\ell$, is associated to the turnover timescale of eddies $\tau_{\ell}$ at this scale. Taking this argument together with dimensional considerations, we can find the relation between the predictability time of the scale $\ell$ (that corresponds to the wavenumber $k_E \sim 1/\ell$). This is $k_E \sim \varepsilon^{1/2} t^{-3/2}$. In the inset of figure \ref{fig:spec_evolution}, we can see that this scaling is found in our data.

We note however, that there is a critical wavenumber $k_c$ below which the evolution of the error spectrum halts, preventing the full decorrelation of the two solutions. The same is observed for $d>3$ for all values of $\Rey$. This critical wavenumber $k_c$ increases with $d$, as shown in figure \ref{fig:spec_vs_d}. For each dimension, the value $k_c$ is found to be approximately constant across all Re. Furthermore, we see that for $d\approx d_c$, there is a strong correlation in the inertial range, whereas only the dissipative scales become decorrelated. For values $d>d_c$, the error decays so there is not a stable final spectrum to do such an analysis. This is suggestive of a connection with the critical $\Rey$ value needed for error growth. If $k_c$ is outside of the dissipative region, error growth can occur, however, as $d$ approaches $d_c$ it appears that $k_c \rightarrow \infty$.

An additional caveat to the location of $k_c$ in our work may be in the choice of eddy damping factor used in the $E_W(k)$ equation. As discussed in section \ref{se:chaosInt}, this choice is arbitrary but does ensure the correlation spectrum is realisable. This choice may also introduce some correlation into the system; however, it is not clear what a more appropriate eddy damping factor would be. It is possible the ideas used in \citep{baerenzung2008spectral} to introduce different damping factors could be employed for the correlation spectrum but we have not pursued this here.
 
\begin{figure}
    \centering
    \includegraphics[width=0.5\textwidth]{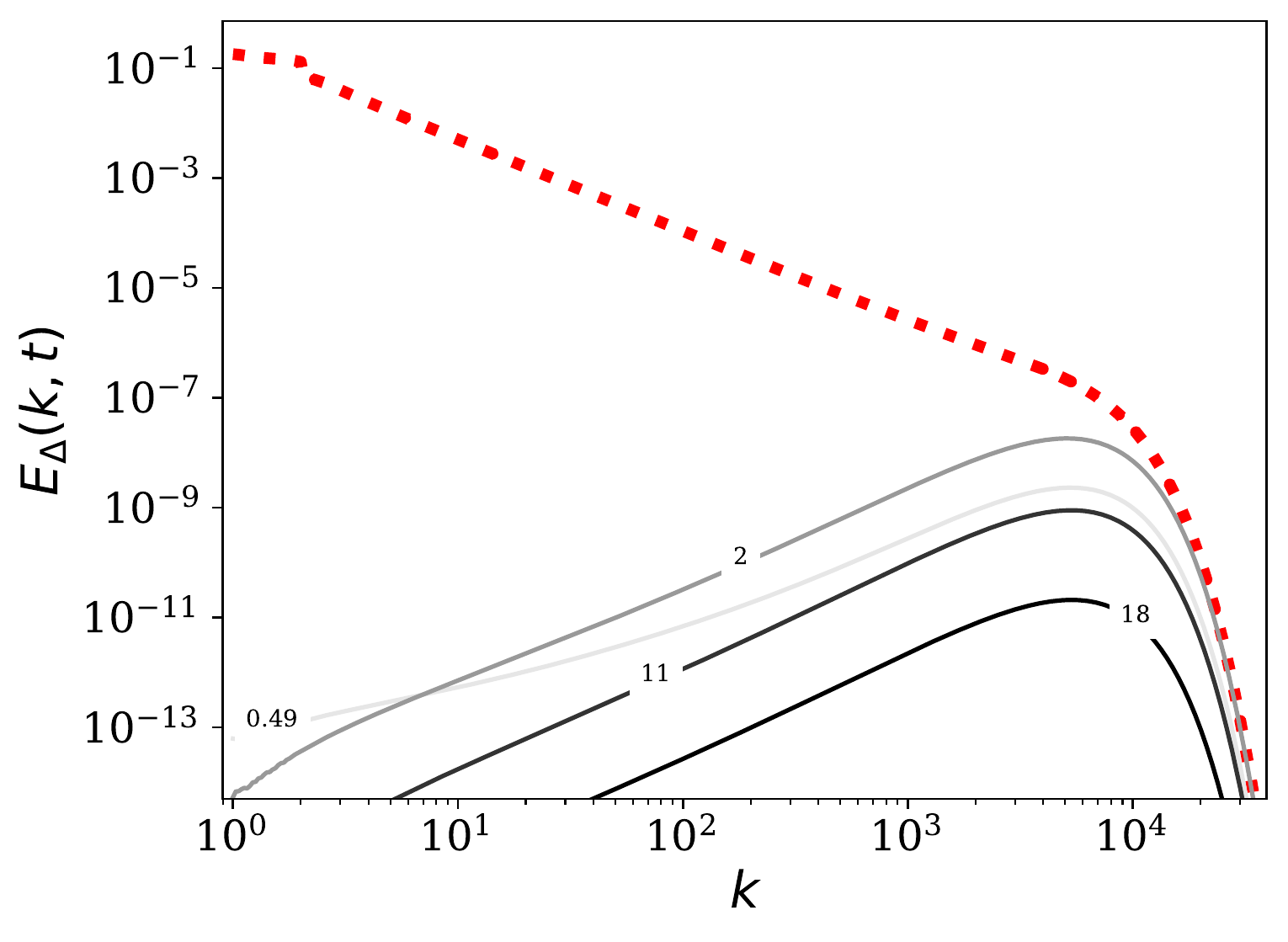}
    \caption{Evolution of $E_{\Delta}(k,t)$, where the labelled solid gray lines correspond to different large eddy turnover times times $t/T = 0.49$, $2$, $11$ and $18$ for a run with $d=6$ and $\nu=1\cdot 10^{-6}$. The red (dotted) line represents the steady state spectrum $E(k)$.}
    \label{fig:spec_evolution_6d}
\end{figure}

The evolution of the spectrum for $d>d_c$ is analysed separately. As seen in figure \ref{fig:spec_evolution_6d}, there is an initial transient where the spectrum grows and changes its form. At this time, $E_{\Delta}(k,t)$ peaks at approximately the Kolmogorov wavenumber $k_{\eta} = \left(\varepsilon/\nu^3\right)^{1/4}$. This corresponds to the initial rapid growth observed in figure \ref{fig:e_growths}. After this time, the error spectrum starts decaying while keeping its form (peaking at $k_{\eta}$), similar to the case of exponential growth. A similar behaviour is observed for $d=7$ and $8$ across the entire range of $\Rey$ explored. Although, as $d$ grows, the initial growth stage is weaker and contrary to $d=6$, the decorrelation is never achieved over the dissipative wavenumbers.   

\subsection{Error Cascades}\label{sec:errorCascade}

Following \citet{leith1972predictability} we consider now the various cascade rates of  (\ref{eq:E_and_Ew}). 
In the above, the terms $T(k)$, $T_{W}(k)$ and $T_{\Delta}(k)$  represent the transfer of energy, correlated energy and decorrelated energy into a wavenumber $k$ respectively. Whilst $T_{X}(k)$ gives the transfer of correlated energy into uncorrelated energy at a given $k$. Focusing on the error equations we note that both $T_W(k,t)$ and  $T_{\Delta}(k)$ are conservative due the antisymmetry of the integrands in $k\leftrightarrow p$. This means that the total rates $\int_0^\infty dk \, T_{W,\Delta}(k) = 0$ , hence, these transfer terms simply transport correlated and uncorrelated energy, respectively, across the different scales, however no net energy is created or removed in these processes. On the other hand, $T_{X}(k,t)$ is non-conservative and we define $\varepsilon_{X} = \int_0^\infty dk \,T_X(k)$ which is the rate of total transfer between correlated and uncorrelated energy. Finally, the correlated and uncorrelated energy dissipation rates are given by $\varepsilon_{W,\Delta} = 2\nu \int_0^\infty k^2 E_{W,\Delta}$. Thus, the net error growth (or decay) is given by \begin{equation}\label{eq:edeltaX}
\partial_t E_{\Delta}(t) = \varepsilon_{\text{net}} \equiv \varepsilon_{X} - \varepsilon_{\Delta} \quad.
\end{equation}

\begin{figure}
 \centering
 \subfigure[]{\label{fig:rates_3d}\includegraphics[width=0.47\linewidth]{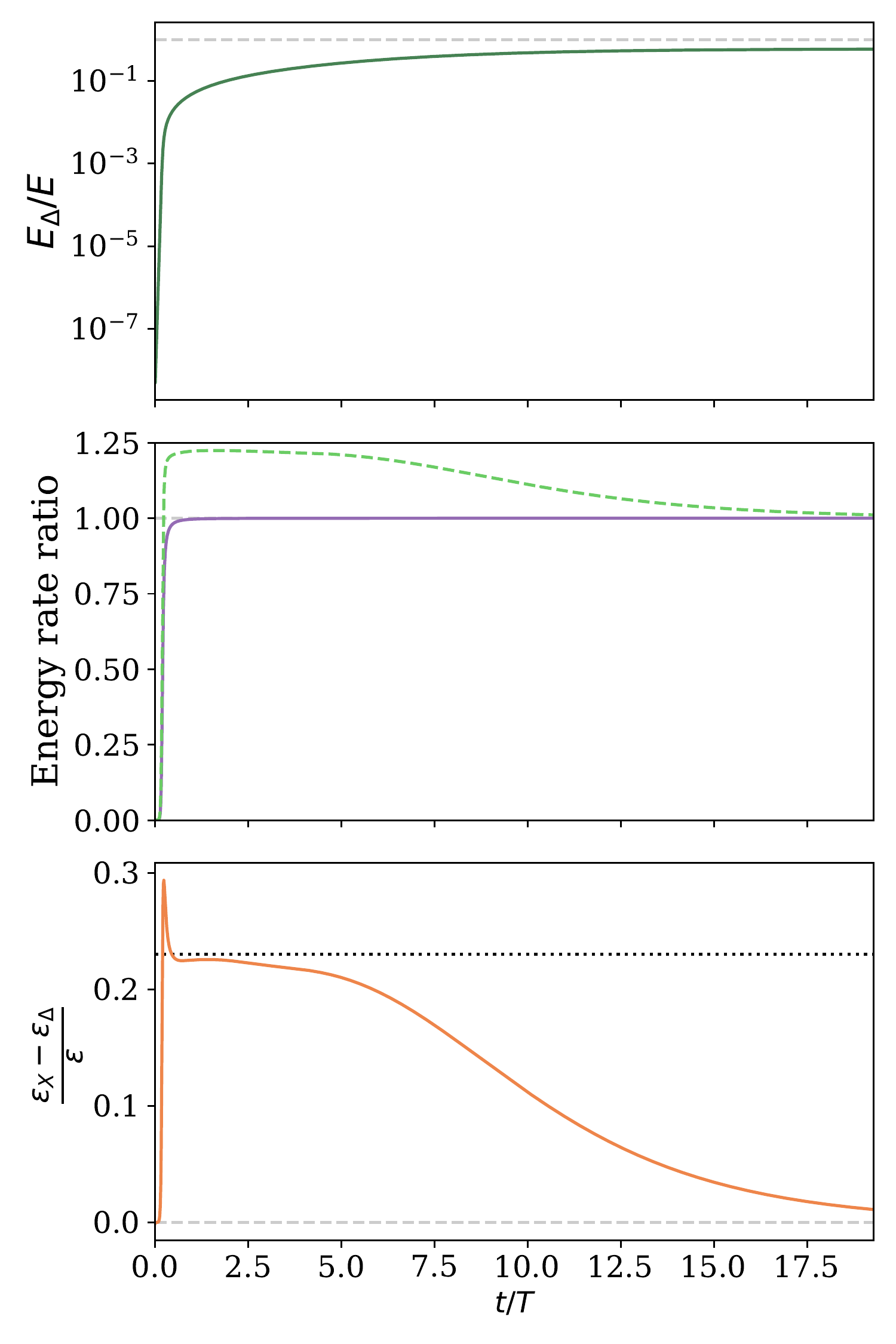}}
 \subfigure[]{\label{fig:rates_4d}\includegraphics[width=0.47\linewidth]{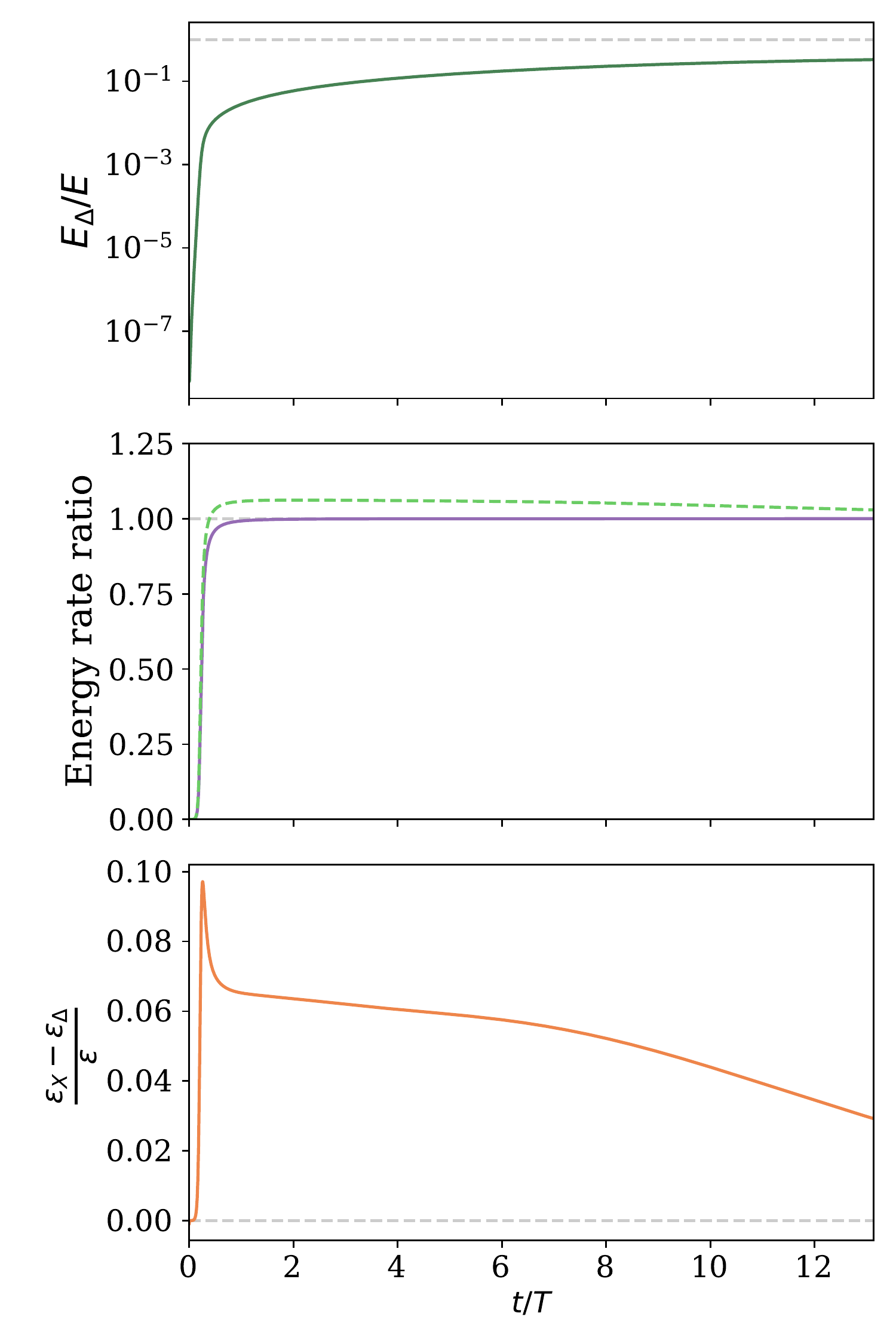}}
 \subfigure[]{\label{fig:rates_5d}\includegraphics[width=0.47\linewidth]{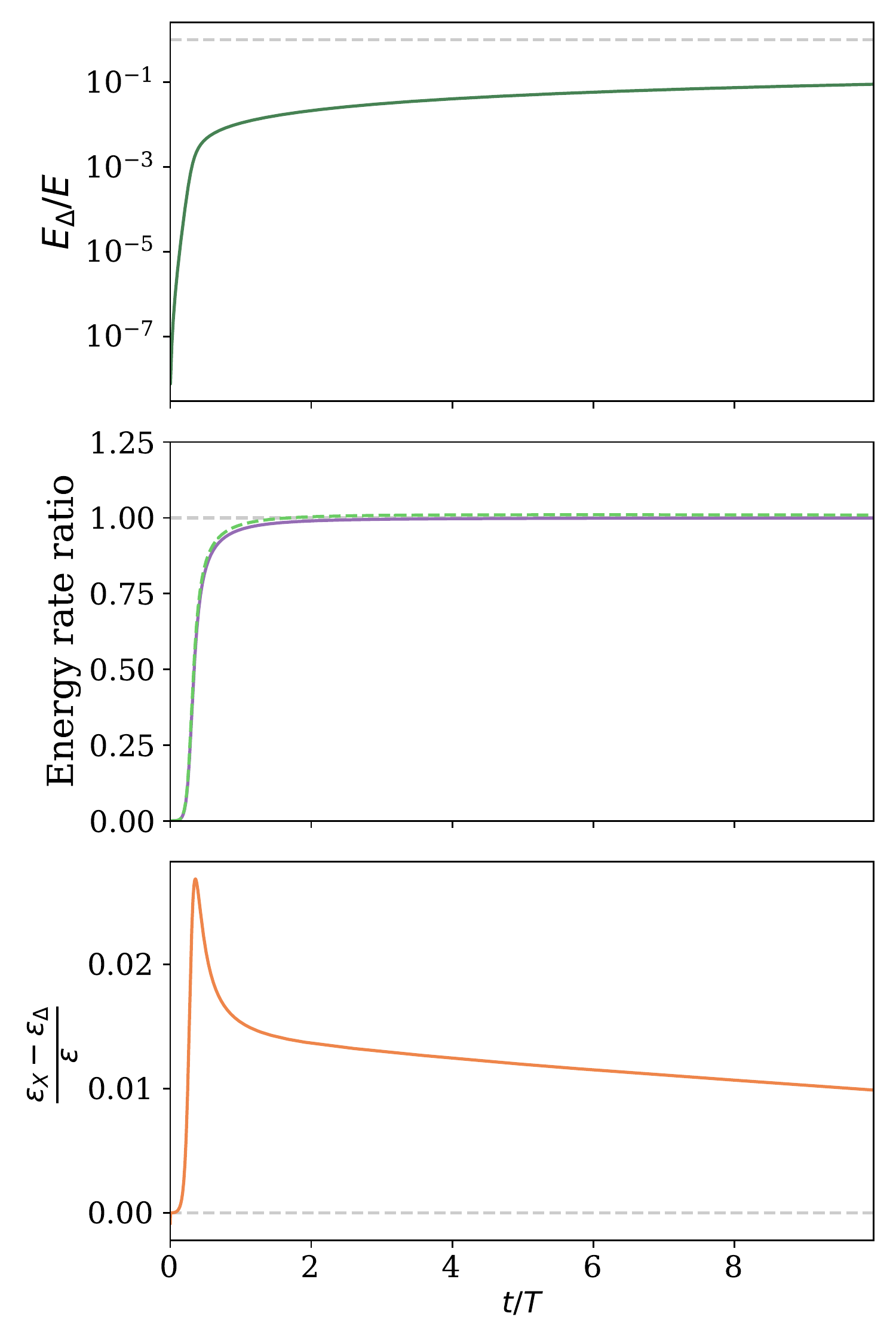}}
 \subfigure[]{\label{fig:rates_6d}\includegraphics[width=0.47\linewidth]{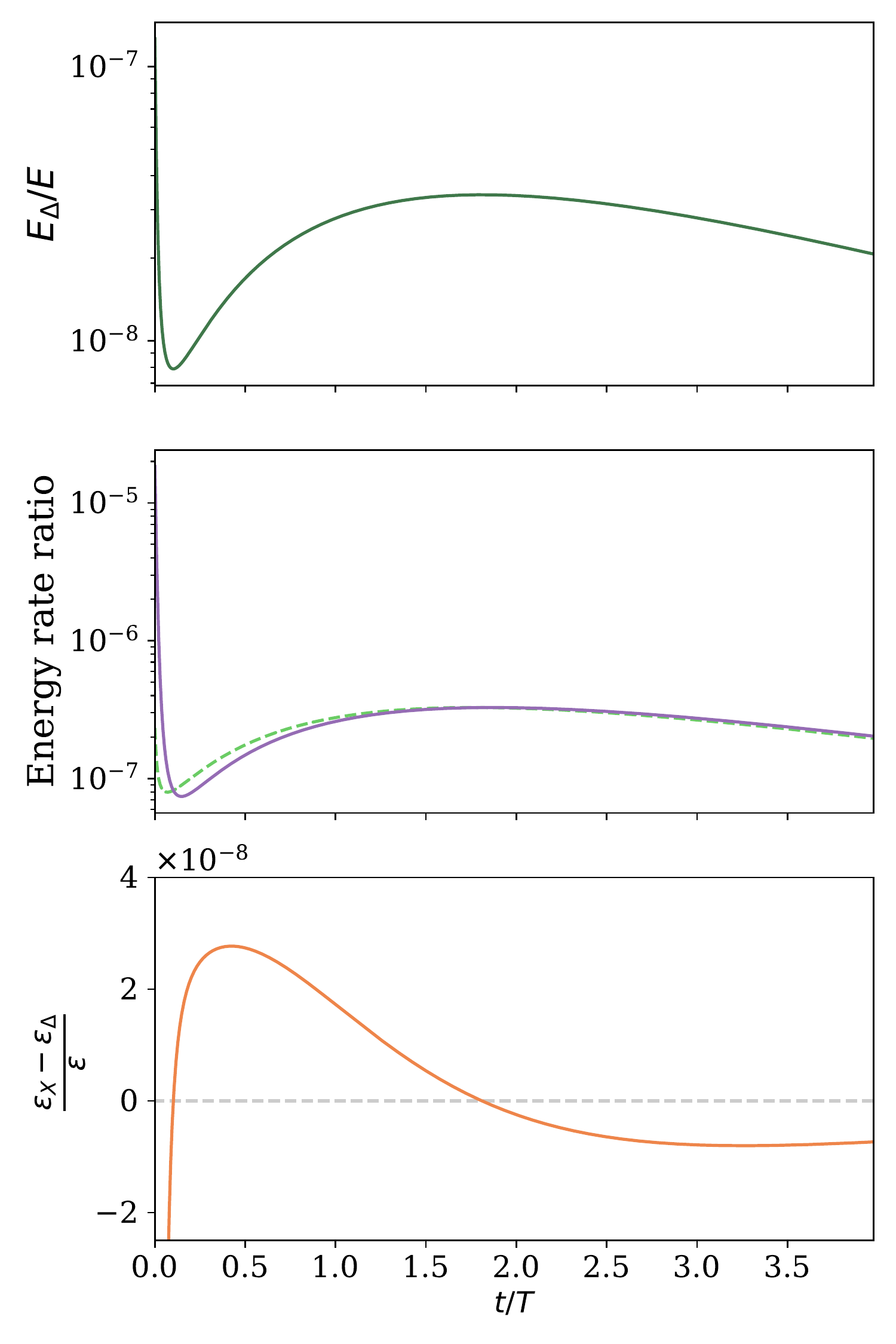}}
\caption{\label{fig:rates_3_to_6} (Top) Evolution of the growth $E_{\Delta}/E$, (Middle) detailed normalised rates $\varepsilon_X/\varepsilon$ (dashed), (Bottom) $\varepsilon_{\Delta}/\varepsilon$ (solid) , and $\varepsilon_{X}/\varepsilon$ (dashed). (a) $d=3$ and $\nu=1.2\cdot 10^{-6}$, (b) $d=4$ and $\nu=1.2\cdot 10^{-6}$, (c) $d=5$ and $\nu=1\cdot 10^{-6}$ and (d) $d=6$ and $\nu=2\cdot 10^{-3}$. For $d=3$, the black dotted line shows Leith and Kraichnan's prediction $\varepsilon_{\text{net}} = 0.23 \varepsilon$ for fully developed turbulence in an infinite Kolmogorov energy spectrum using the TFM \citep{leith1972predictability}.} 
\end{figure} 

\begin{figure}
    \centering
    \includegraphics[width =0.7\textwidth]{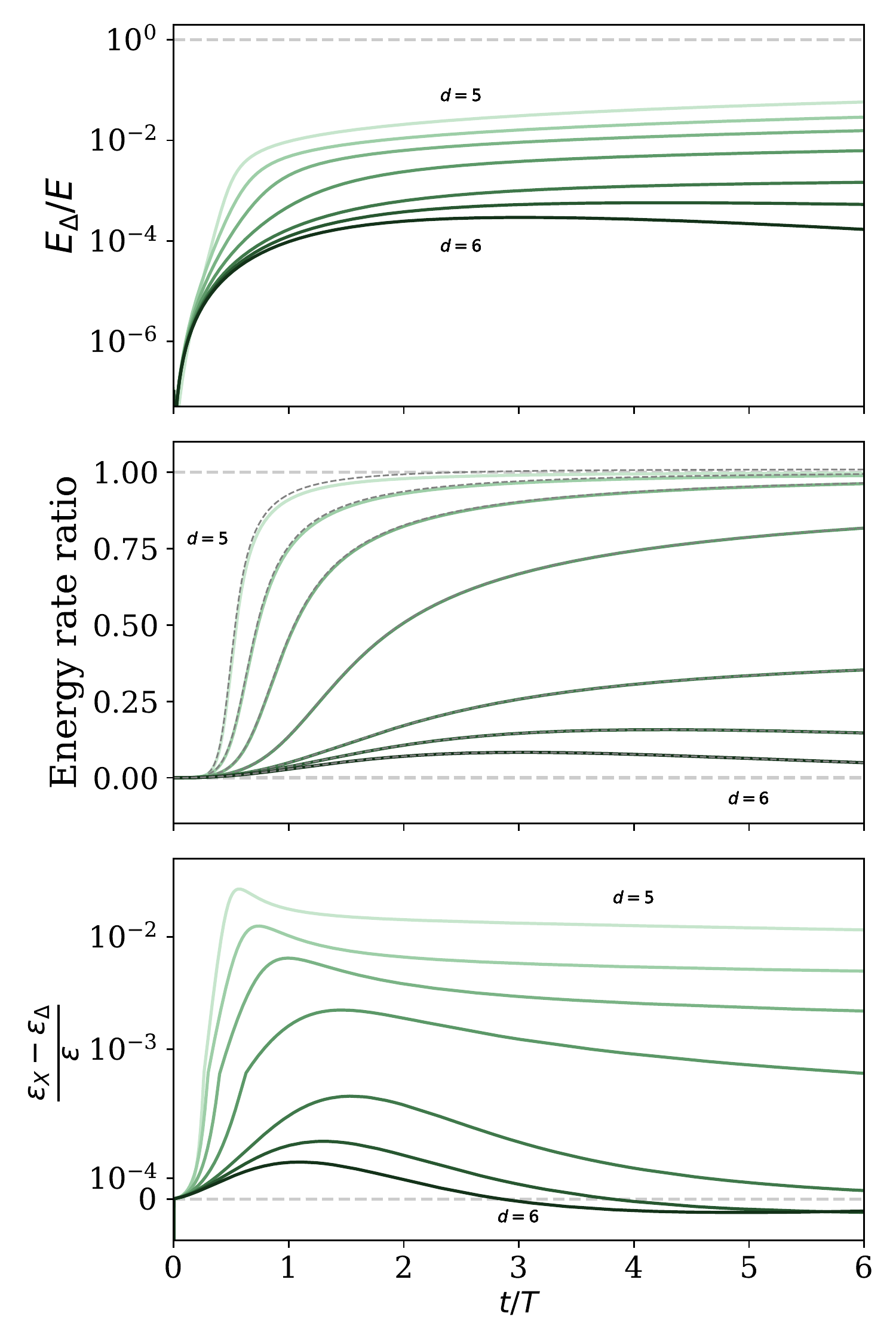}
    \caption{(Top) Evolution of the growth $E_{\Delta}/E$ , (Middle) detailed normalised rates $\varepsilon_X/\varepsilon$ (gray dotted) and $\varepsilon_{\Delta}/\varepsilon$ (green solid), (Bottom) $\varepsilon_{\text{net}}/\varepsilon$ with a symmetrical logarithmic scale in the vertical axis, for dimensions $d=5$, $5.3$, $5.5$, $5.7$, $5.85$, $5.9$, $5.95$ and $6$ and $\nu=2\cdot 10^{-6}$. In all cases the line colour becomes darker as we move from 5 to 6 dimensions.}
    \label{fig:rates_nonintd}
\end{figure} 

We consider the evolution of these cascade rates in time. In figure \ref{fig:rates_3_to_6} we show the evolution of the normalised uncorrelated energy $E_{\Delta}/E$ together with the detailed rates $\varepsilon_{X,\Delta}$ and the net rate $\varepsilon_{\text{net}}$ for different dimensions. For $d=3,4$ and $5$, we observe the existence of a linear growth that follows the exponential growth, and it corresponds to the growth of the self-similar spectrum observed in figure \ref{fig:spec_evolution}. The linear stage is noticeable for the largest $\Rey$ runs, where there is a larger inertial range. For lower $\Rey$ runs, the exponential growth is followed by the saturation stage. After the linear stage, the rate decays until saturation. We see in the detailed plots, that the $\varepsilon_{\Delta}/\varepsilon$ grows and saturates at 1 right after the exponential growth stage, much faster than the saturation of $E_{\Delta}/E$. At the same time, $\varepsilon_{X}/\varepsilon$ grows and saturates also to a value slightly greater than 1, but then starts to decrease as the uncorrelated energy reaches its saturation value. The net rate shows an initial peak that coincides with the end of the exponential growth that is common to all runs for dimensions $d\leq5$. For $d=3$ we see that the value of the net rate agrees with Leith and Kraichnan's prediction, $\varepsilon_{\text{net}}=0.23\varepsilon$ \citep{leith1972predictability} which was computed as an approximation to the infinite inertial range case by fixing $E(k)$ in time as a Kolmogorov spectrum.

For $d=6$, we note that during the initial stage there is a competition between $\varepsilon_{X}$ and $\varepsilon_{\Delta}$, that results in a rapid decay of the uncorrelated energy that corresponds to the stage where $E_{\Delta}(k)$ adopts its typical form as seen in figure \ref{fig:spec_evolution_6d}. This is followed by a short period of growth, until the net rate becomes negative again resulting in the final decay of $E_{\Delta}$. Additionally, we show the behaviour of the net rate during transition to non-chaotic regime in detail in figure \ref{fig:rates_nonintd}. For $d>d_c$, we observe the typical evolution of the net rate with a positive value resulting in a growth stage followed by an inversion of the sign that results in the final decay. There are two interesting remarks for the behaviour of the net rate for values of $5<d<d_c$. First, we note that it takes longer for the dissipation rate of uncorrelated energy $\varepsilon_{\Delta}$ to reach saturation as $d$ increases. At the same time, we see that the sharp peak that is typically observed for $d\leq 5$ and that corresponds to the initial exponential growth stage, starts to soften gradually as $d$ approaches $d_c$.

\begin{figure}
 \centering
 \subfigure[]{\label{fig:flux3d}\includegraphics[width=0.32\linewidth]{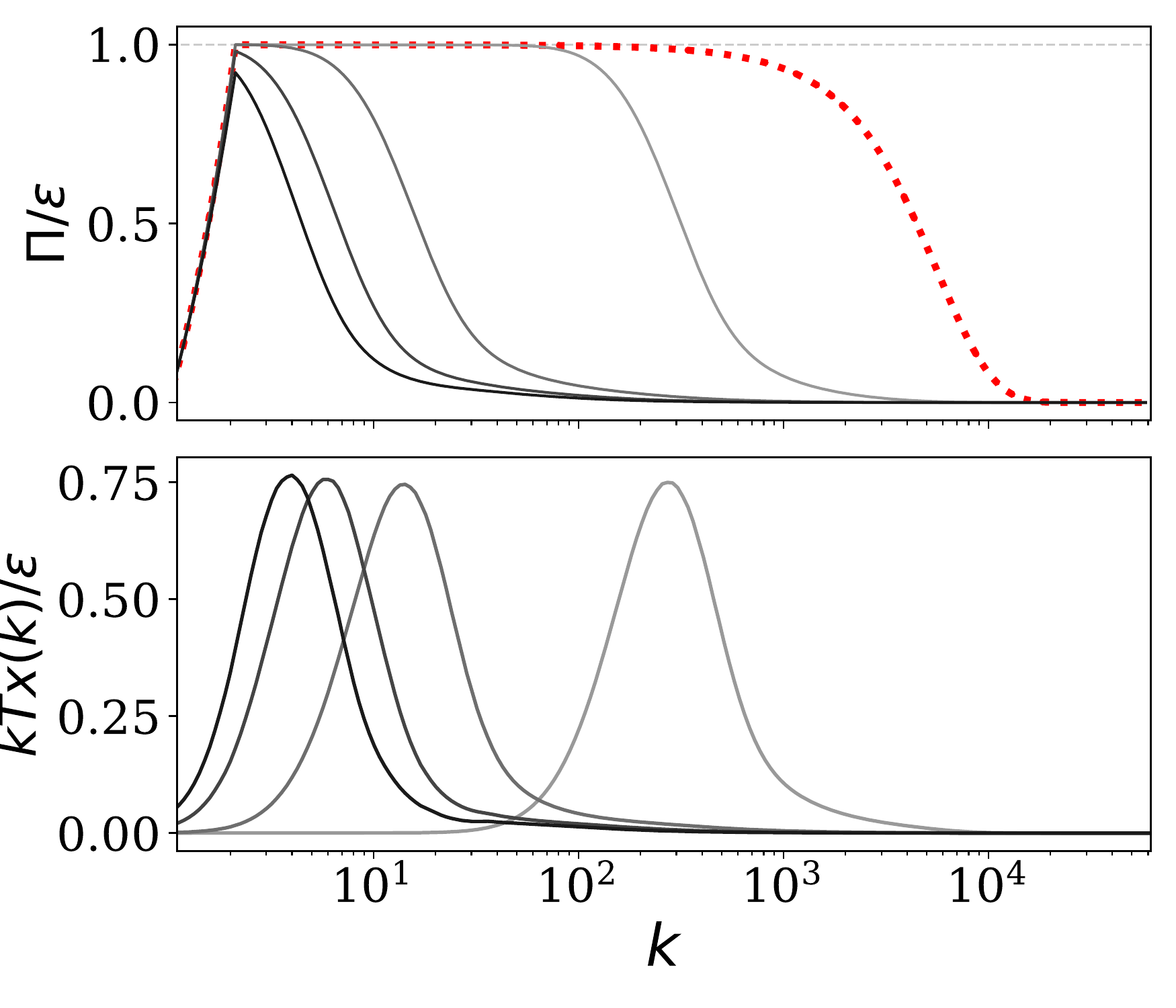}}
 \subfigure[]{\label{fig:flux4d}\includegraphics[width=0.32\linewidth]{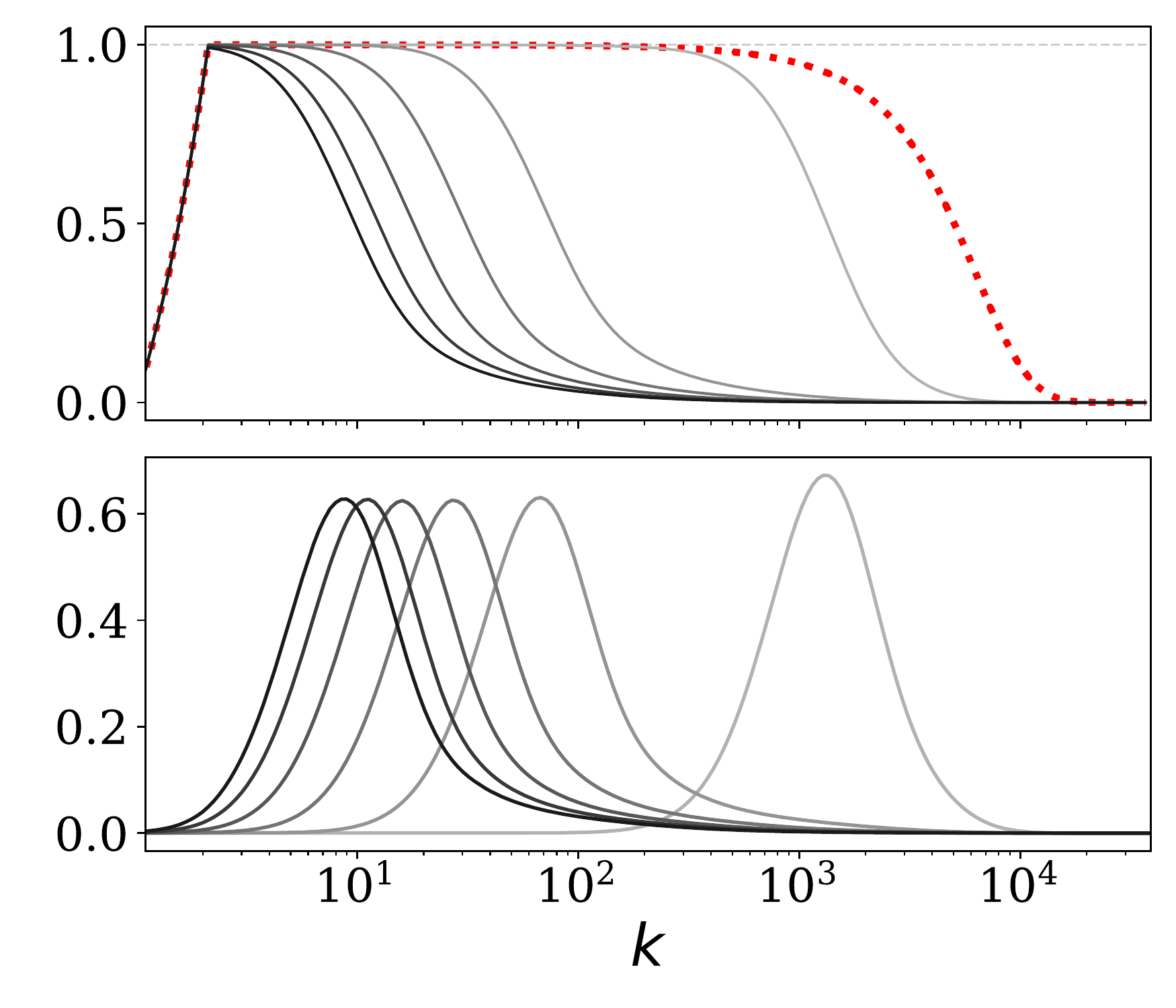}}
 \subfigure[]{\label{fig:flux5d}\includegraphics[width=0.32\linewidth]{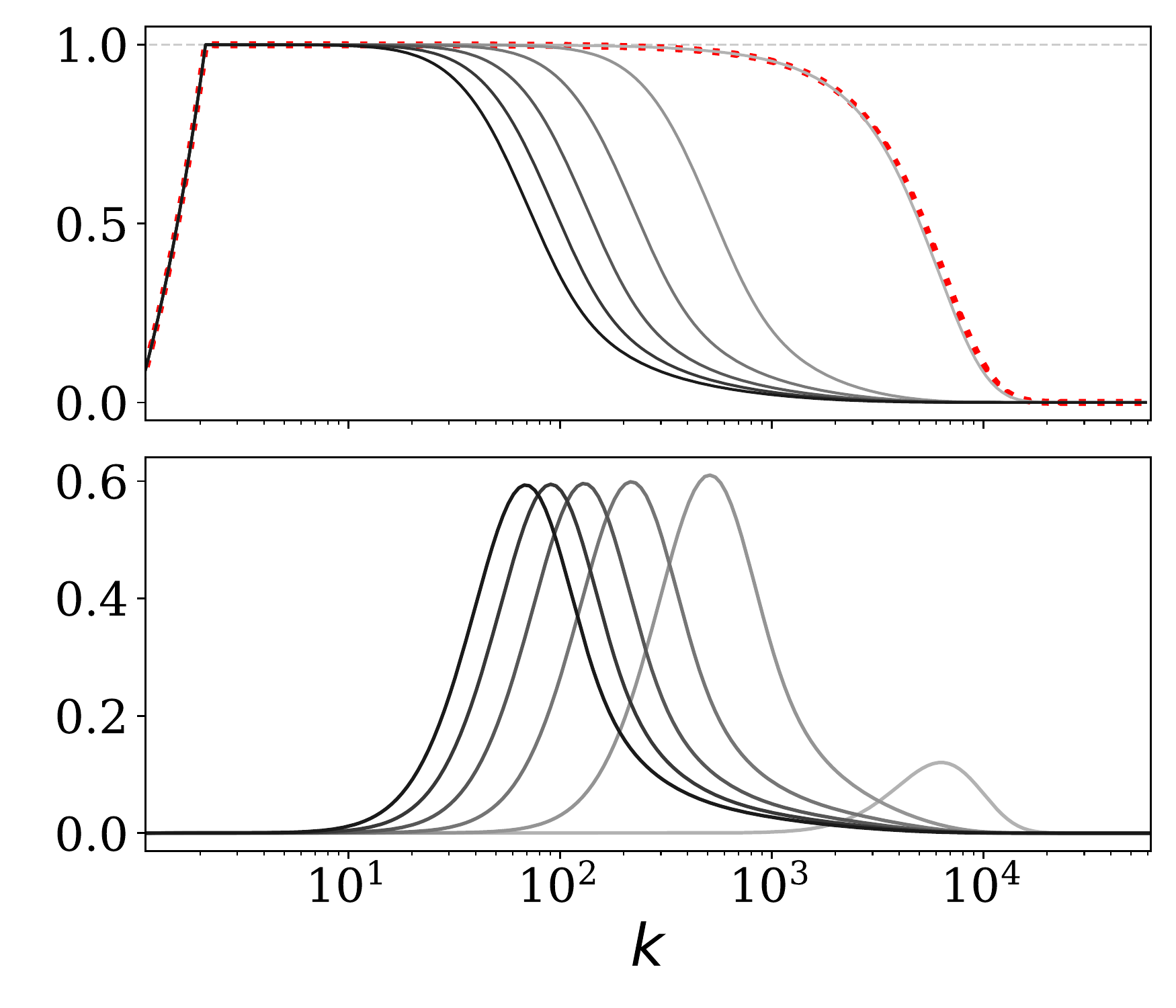}}
\caption{\label{fig:flux_3_to_5} (Top) Evolution of the of the correlated energy flux $\Pi_W(k)$  corresponding to the correlated energy cascade (solid lines), along with the energy flux $\Pi_u(k)$ corresponding to the energy cascade (dotted), and (Bottom) compensated transfer spectrum $k\,T_X(k)/\varepsilon$. Time evolution goes from light to dark lines. The viscosity for all runs is $\nu=2 \cdot 10^{-6}$. (a) $d=3$ and $t/T =0.5$, $5$, $7$ and $10$, (b) $d=4$ and $t/T=0.5$, $2.5$, $5$, $7$, $9$ and $11$, (c) $d=5$ and $t/T=0.3$, $2$, $4$, $5$, $7$ and $9$} 
\end{figure}

\begin{figure}
    \centering
    \includegraphics[width =0.7\textwidth]{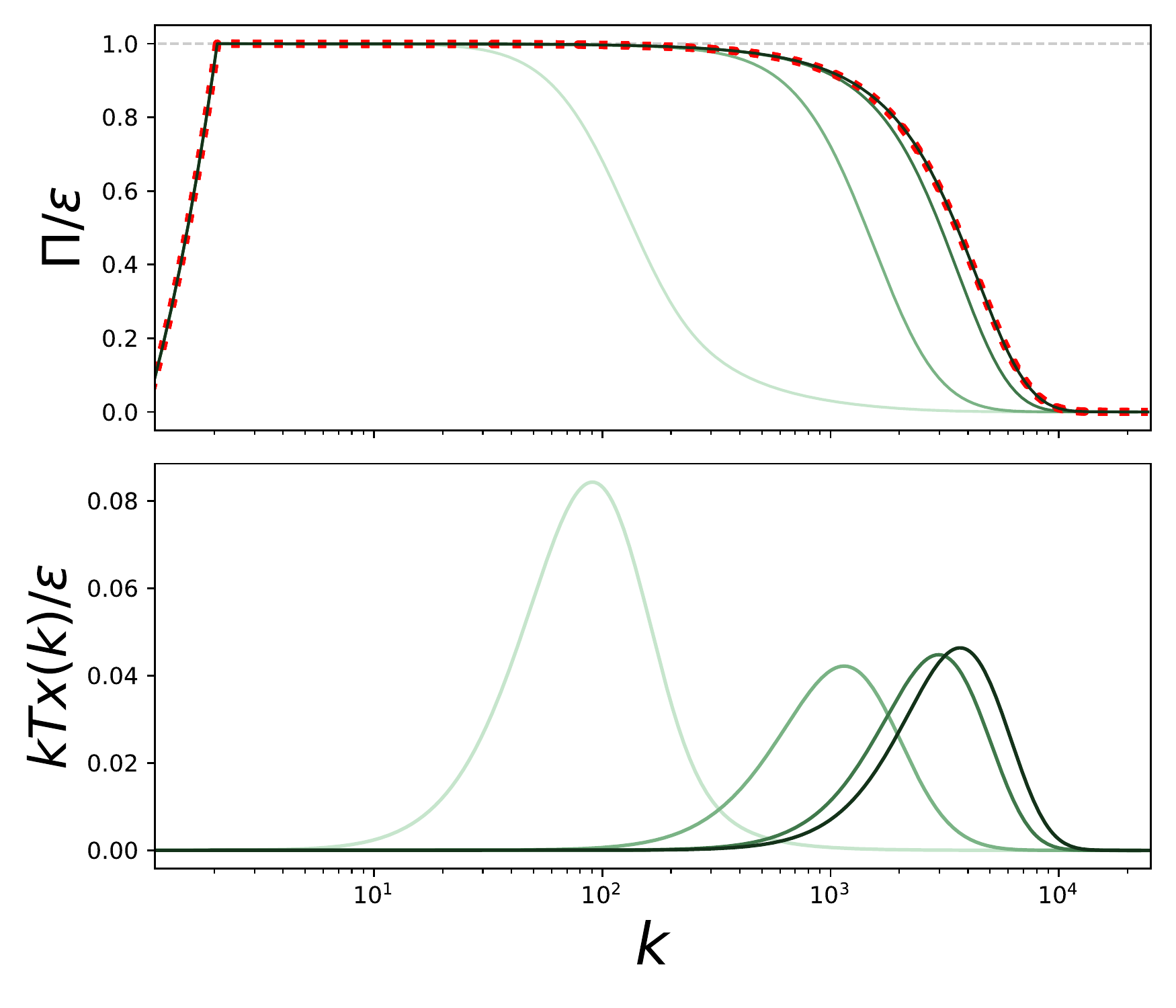}
    \caption{(Top) Correlated energy flux $\Pi_W(k)$ corresponding to the correlated energy cascade (solid green lines), along with the energy flux $\Pi_u(k)$ corresponding to the energy cascade (red dotted), and (Bottom) compensated transfer spectrum $k\,T_X(k)/\varepsilon$. The viscosity for all runs is $\nu=2 \cdot 10^{-6}$ and dimensions are $d = 5.3$ (lighter green), $5.7$, $5.85$ and $6$ (darker green).}
    \label{fig:nonintd_flux}
\end{figure}

From the above analysis it is clear there is a competition between the generation of uncorrelated energy and its sweeping out by the cascade. To show more explicitly the interplay between uncorrelated and correlated energy we look at the evolution of the flux of correlated energy 
\begin{equation}
\Pi_W(k) = - \int_0^k dp \, T_W(p) = \int_k^{\infty} dp \, T_W(p) \quad ,
\end{equation}
that represents the forward cascade of correlated energy compared to the energy flux \begin{equation}
\Pi_u(k) = - \int_0^k dp \, T(k) = \int_k^{\infty} dp \, T(k) \quad ,
\end{equation} together with the evolution of the transfer from correlated to uncorrelated energy $T_X(k)$. We observe in figure \ref{fig:flux_3_to_5} that the cascade of correlated energy towards higher wavenumbers becomes more effective as the dimension grows. In figure \ref{fig:flux3d} it is clear that the correlated energy is transferred forward from the forcing range towards a wavenumber where it is then transferred (as observed in the $T_X$ plot) from correlated to uncorrelated energy, which is then swept out by the uncorrelated energy flux $\Pi_{\Delta} = \Pi_u - \Pi_W$, which is not shown in the plot. Initially, the transfer from correlated to uncorrelated energy takes place at the dissipation wavenumber, but as the system evolves, this transfer decreases towards $k_c$. For $d=3$, the value of $k_c$ is at the very forcing range, whereas for $d=5$ the forward cascade is more effective and the transfer to uncorrelated energy occurs at the inertial range. Finally, in figure \ref{fig:nonintd_flux} we show the flux transition from $d=5.3$ to $d=6$. There we observe that as dimension increases, the transfer from correlated to uncorrelated occurs at the dissipation range, and the cascade of energy and correlated energy are equal for the whole inertial range. As such, by considering the correlated energy flux we find the the value $k_c$ is determined by this flux.

It is clear that below the critical dimension and at $\Rey > \Rey_c$ the production of uncorrelated energy wins this competition and the error between the two velocity fields grows. However, at the critical dimension the situation is altered and the removal of uncorrelated energy by the cascade begins to dominate. This is interesting in light of the results we found in \citep{clark2021effect}, most notably the finding of a maximum enstrophy production near five dimensions. The roles played by vortex stretching and strain self-amplification, and thus also enstrophy production, in the three-dimensional turbulent energy cascade have been the subject of numerous recent investigations \citep{carbone2020vortex, johnson2020energy, buaria2020vortex, bos2021three}. Indeed, \citet{bos2021three} demonstrated that removing both vortex stretching and strain self-amplification from the EDQNM closure in three dimensions results in enstrophy conservation and an inertial range with exponent -3. However, removing this term also acts to suppress strain self-amplification so a distinction between the importance of vortex stretching versus strain self-amplification could not be made. In \citet{carbone2020vortex} the two effects were distinguished, and strain self-amplification was identified as having the dominant effect on the energy cascade. Nevertheless, in \citep{clark2021effect} the dimension of maximum enstrophy production will correspond to the case where the combined influence of both effects on the production enstrophy are maximised.

The proposed role of vortex stretching and strain self-amplification in error growth and the production of uncorrelated energy was also considered in \citep{clark2020chaotic}. Here the transition between two- and three-dimensional dynamics in a thin-layer was studied via the maximal Lyapunov exponent. A discontinuous transition in the scaling of the exponent was found as the flow moved from two- to three-dimensional dynamics. On the two-dimensional side of the transition there is neither vortex stretching nor strain self-amplification and it is found that the Lyapunov exponent is determined by the injection of enstrophy. On the three-dimensional side, as in this work, the exponents are found to depend on the Reynolds number. The discontinuous nature of the transition suggests that these two processes are intimately related to error growth. Notably, two dimensional turbulence is still chaotic without either process, hence disentangling the individual roles of each process in the growth of error may be challenging.  

\section{Conclusions}
\label{se:conclusions}

Motivated by the idea of critical dimension for fluid turbulence above which K41 becomes exact we have performed a numerical study of the evolution of the error $E_{\Delta}(k,t)$ for homogeneous and isotropic $d$-dimensional turbulence. We consider spatial dimensions ranging from $3$ to $8$, including non-integer values, using an EDQNM closure approximation. Our work has focused on analysing growth of $\Delta(t)$ through the maximal Lyapunov exponent and the production and removal of uncorrelated energy.

A critical dimension for error growth of $d_c\approx 5.88$ is found, above which the error $\Delta(t)$ decays instead of growing for large times. This value is close to the dimension of maximal enstrophy production found in \citep{clark2021effect}. It is proposed that these dimensions are the related and thus that enstrophy production and error growth in turbulence are linked. As the critical dimension is approached the competition between the generation of uncorrelated energy, and thus error, and the sweeping out of error by the energy cascade begins to move in favour of the cascade. We suggest this is related to a change in the relative strengths of strain self-amplification and vortex stretching. These two processes are directly related to the energy cascade \citep{carbone2020vortex, johnson2020energy, buaria2020vortex, bos2021three} and to the production term in the enstrophy equation.

Considering the discussion of cascade fluctuations and their role in determining the validity of K41 made by \citet{kraichnan_1974, kraichnan1991turbulent} the possibility of a reduction in the importance of vortex stretching beyond $d_c$ is interesting. If a connection between vortex stretching and error growth can be made, our work may suggest a suppression of fluctuations beyond $d_c$. This, of course, can only be entirely speculative due to the use of a closure approximation in both this work and \citep{clark2021effect}. Still, it is a property that appears consistently in the numerical calculations of the EDQNM equations we have used. The EDQNM closure approximation is widely used in many theoretical and numerical works so even if this property is not present in DNS or real flows, it is still an interesting property to look at in the future using other closure approximations, or different numerical realizations of the same equations. The trend observed of decreasing $\lambda$ when going from $d=3$ to $d=4$, certainly sets the question for higher dimensions, but that remains for future analysis. Studying the error growth in $d$-dimensions using other closure models could discard or support this finding. Additionally, an interesting direction for further study, is to look at DNS at small box sizes in $d=6$ to see if the error growth is present or not. Although the range of $\Rey$ that can be explored in DNS is very small, these simulations might shed some light on the existence of the non-chaotic regime.

Another interesting angle to view these results from is the possibility of Burgers equation dynamics at high spatial dimension. As discussed earlier in this work, this idea has been considered in previous works \citep{fournier1978infinite, gotoh2007statistical, falkovich2010new}. Due to the incompressibility condition being the sum of $d$ terms as $d$ is increased the effect of the pressure term on each individual velocity component will be diminished. The Burgers equation is given by the Navier-Stokes equation with the pressure gradient term removed, hence the suggestions that the dynamics of high dimensional turbulence may resemble those of the Burgers equation. Our results provide another possible connection in this direction as it is known that the Burgers equation is integrable and thus does not exhibit chaos \citep{hopf1950partial, cole1951quasi}. It may be that the transition to a non-chaotic regime is related to the role of pressure, however this will require further investigation. This is the opposite picture to what was described in the previous paragraph. If the Navier-Stokes equations in high dimension exhibit Burgers equations statistics this will involve an extreme anomalous scaling.

In summary, using the EDQNM approximation for isotropic turbulence we have found a critical dimension for error growth at $d_c \approx 5.88$. By considering previous results for $d$ dimension EDQNM calculations we propose this dimension is related the dimension of maximal enstrophy production found in \citep{clark2021effect}. By considering the competition between the energy cascade and error production we relate our findings to fluctuations in the energy cascade and the idea of a critical dimension above which K41 is exact. Additionally, we consider the possibility of a Burgers equation statistics limit for the Navier-Stokes equation in high spatial dimension where an extreme anomalous scaling contrary to K41 could be expected. Determining which of these scenarios is correct will require further study of turbulence in spatial dimensions above three.

\section*{Acknowledgements}
This work has used resources from ARCHER via the Director’s Time
budget. D. C. and A. A. are supported by the University of Edinburgh. A.B. acknowledges partial funding from the U.K. Science and Technology Facilities Council.

\section*{Declaration of Interests}
The authors report no conflict of interest

\bibliography{sample.bib}
\bibliographystyle{jfm}

\end{document}